\documentclass[aps, reprint, amsmath, amssymb, superscriptaddress, prb,longbibliography, preprintnumbers]{revtex4-2}

\usepackage[hidelinks]{hyperref} 
\usepackage[dvipsnames]{xcolor}
\usepackage{colortbl}
\hypersetup{
    colorlinks,
    linkcolor={red!80!black},
    citecolor={blue!50!black},
    urlcolor={magenta!100!black}
}
\usepackage{graphicx}
\usepackage{dcolumn}
\usepackage{bm}
\usepackage{mathrsfs}
\usepackage{verbatim}
\usepackage{soul} 
\usepackage{multirow}
\usepackage{braket}

\usepackage{changes}

\usepackage{lineno}

\begin{document}

\preprint{LA-UR-25-29854}

\title{Mixed Stochastic-Deterministic Density Functional Theoretic Decomposition of Kubo-Greenwood Conductivities in the Projector Augmented Wave Formalism}
\author{Vidushi Sharma}
\email{vidushi@princeton.edu}
\affiliation{Theoretical Division, Los Alamos National Laboratory, Los Alamos, New Mexico 87545, USA}
\affiliation{Center for Nonlinear Studies (CNLS), Los Alamos National Laboratory, Los Alamos, New Mexico 87545, USA}
\affiliation{Applied Materials and Sustainability Sciences, Princeton Plasma Physics Laboratory, Princeton, NJ 08540-6655, USA}
\author{Lee A. Collins}
\affiliation{Theoretical Division, Los Alamos National Laboratory, Los Alamos, New Mexico 87545, USA}
\author{Alexander J. White}
\email{alwhite@lanl.gov}
\affiliation{Theoretical Division, Los Alamos National Laboratory, Los Alamos, New Mexico 87545, USA}
\date{\today}

\begin{abstract}
Pairing the accuracy of Kohn-Sham density-functional framework with the efficiency of a stochastic algorithmic approach, mixed stochastic-deterministic Density Functional Theory (mDFT) achieves a favorable computational scaling with system sizes and electronic temperatures.
We employ the recently developed mDFT formalism to investigate the dynamic charge-transport properties of systems in the warm dense matter regime.
The optical conductivity spectra are computed for single- and multi- component mixtures of carbon, hydrogen, and beryllium using two complementary approaches: Kubo-Greenwood in the mDFT picture and real-time Time-Dependent mDFT.
We further devise a decomposition of the Onsager coefficients leading up to the Kubo-Greenwood spectra to exhibit contributions from the deterministic, stochastic, and mixed electronic state transitions at different incident photon energies.
\end{abstract}

                              
\maketitle

\section{Introduction}


A first-principles investigation of equations of state and charge-transport properties lies at the forefront of Warm Dense Matter (WDM) research, and constitutes a rich landscape of open questions yet to be explored \cite{Falk2018, 2025roadmapwarmdensematter}. WDM systems of astrophysical interest include the interiors of icy giant gas planets, and compact objects such as brown dwarfs and white dwarf envelopes \cite{Militzer16, Helled2020, SAUMON20221, Doris23, Roy24, Roy2025}. WDM has also been hypothesized to exist in icy moons of outer planets, which might harbor liquid water seas with possible applications to the current Europa Clipper mission \cite{Europa}.
In inertial confinement fusion, the compression path of the deuterium–tritium fuel crosses the WDM regime, making charge-transport properties such as resistivity, conductivity, electron-ion coupling, and opacities essential for understanding instabilities and achieving ignition \cite{ICF2020, ICF2021, Grabowski20,Stanek24,Hurricane23, Hu24, Haines24, Allen2025,TWhite2025}. The generation and characterization of WDM in the form of laboratory plasmas by bright X-ray sources continues to push the field forward \cite{Dorchies15, Preston17, Falk2018, Mahieu2018, Frydrych2020, Chen21, Doppner2023, Mercadier2024, Bespalov25, Cordova25}.

Most \textit{ab initio} approaches treat WDM as a disordered liquid (metal) at temperatures of the order of Fermi energy, that is, a few to hundreds of eV's. In this regime, optical conductivity and related electron-transport properties are typically described using the Kubo–Greenwood model \cite{Dufty2018, White2024, ALLEN2006165}.
The Kubo–Greenwood (KG) formula derives from a broader class of Kubo formulas for two-point quantum correlation functions \cite{Kubo1957, Greenwood1958}, and is based on transitions between filled and empty band states, implying that in practice, one requires knowledge of a large number of eigenstates (bands) up to an energy cutoff, thus making the operation largely memory-intensive for most simulated system sizes.

The computational complexity of traditional Kohn-Sham Density Functional Theory (KS-DFT) scales as $\mathcal{O}(N^3T^3)$, in contrast to the $\mathcal{O}(NT^{-1})$ scaling of stochastic DFT, where $N$ denotes the number of particles (i.e., electrons), and $T$ is the temperature. Building on this principle, our previous works have demonstrated the utility of the mixed stochastic–deterministic DFT (mDFT) framework for studying physical systems over a wide distribution of temperatures and densities -- leveraging the deterministic component at low-$T$ and shifting toward predominantly stochastic contributions at high-$T$ \cite{White2020, Sharma2023, Sharma2025}. Another advantage of the mDFT approach is its ability to improve the accuracy and precision of any physical observable, over purely stochastic DFT, for a given computational cost. This amounts to a reliable prediction of material properties in extreme states of matter. 

In this work, we employ mDFT as the electronic structure method in conjunction with the KG model to derive charge-transport coefficients constituting the electron-driven electrical and thermal conductivities. We previously utilized this approach to calculate electrical and thermal conductivities for the second charged-particle transport coefficient code comparison workshop, which was held in Livermore, California, on 24–27 July 2023 \cite{Stanek24}. Here, we present the formalism and workflow for conductivity calculations within the mDFT framework, and further unpack the contributions of stochastic, KS, and mixed subspace transitions to the full Onsager coefficients and KG conductivity. Since KG is a fixed-orbital approximation to linear-response theory without real-time electron dynamics, we also evaluate conductivity as an optical response from an alternate perspective, namely, real-time Time-Dependent Density Functional Theory (TDDFT), employing the time evolution of mixed stochastic-deterministic states.

The paper is organized as follows. Section \ref{sec:transport} begins with an overview of the computational details and the two electronic structure methods used to compute the electron transport properties; Sections \ref{sec:KG} and   \ref{sec:tddft} cover the Kubo-Greenwood and mixed stochastic-deterministic TDDFT formalisms, respectively, as pertinent to WDM, and Section \ref{sec:paw-tddft} details the PAW formalism employed in much of this work. Physical systems, their charge-transport (Onsager) coefficients, and AC conductivity spectra are discussed in Section \ref{sec:results}. Finally, Section \ref{sec:conclusion} outlines the conclusions of this study.

\section{\label{sec:transport} Charge Transport in finite-temperature DFT}
\begin{table*}
    \centering
    \caption{\label{tab:tab1} Electrical and (electronic
) thermal DC conductivities obtained with the Kubo-Greenwood formalism in KS-DFT ($\sigma^{\psi}$, $\kappa^{\psi}$ resp.) and mixed DFT ($\sigma^{\psi\chi}$, $\kappa^{\psi\chi}$ resp.); electrical DC conductivity from real-time Time-Dependent mixed DFT ($\sigma^{\text{m-TDDFT}}$) for systems at given mass density $(\rho)$ and temperature ($T$). The DC values $(\omega\rightarrow 0)$ are averaged over ten uncorrelated snapshots obtained from an equilibrated thermostatted MD trajectory.}
    \begin{tabular}{ccc|ccc|ccc|c} \hline\hline
         System & $\rho$ [g/cm$^3$] & $T$ [eV] & $N_\psi$ & $\sigma^{\psi}_{\omega\rightarrow0}$ [MS/m] & $\kappa^{\psi}_{\omega\rightarrow0}$ [W/m-K] & $N_\psi/N_\chi$ & $\sigma^{\psi\chi}_{\omega\rightarrow0}$ [MS/m] &  $\kappa^{\psi\chi}_{\omega\rightarrow0}$ [W/m-K]  & $\sigma^{\text{m-TDDFT}}_{\omega\rightarrow0}$ [MS/m]\\ \hline
         CH & $0.90$ & $7.8$ & 2560 & $0.22 \pm 0.01$ & $697.56 \pm 68.78$ & 768/64 & $0.22 \pm 0.01$ & $690.53 \pm 68.70$ & $0.22 \pm 0.01$\\
         Be & $1.84$ & $4.4$ & $2560$ & $0.63 \pm 0.01$ & $689.96 \pm 11.28$ & $256/64$ & $0.62 \pm 0.04$ & $689.70 \pm 84.45$ & $0.64 \pm 0.10$\\
         CH/Be & $1.37$ & $5.0$ & $3456$ & $0.28 \pm 0.01$ & $450.38 \pm 6.96$ & $1024/64$ & $0.28 \pm 0.01$ & $484.59 \pm 39.07$ & $0.27 \pm 0.04$\\ \hline
    \end{tabular}
\end{table*}
\paragraph{Methodology.} 
The system setup at a given density and temperature $(\rho, T)$ conditions consists of $\text{CH}: 64/64$, $\text{Be}: 128$, $\text{CH/Be}: 128/128/128$ atoms, respectively.
We employ Projector Augmented Wave (PAW) potentials \cite{Blochl1994, Kresse1999} to self-consistently solve for the semi-core electrons of hydrogen $(1s^1)$, beryllium $(1s^2,\,2s^2)$, and carbon $(2s^2,\,2p^2)$ atoms. A converged planewave kinetic energy cutoff is chosen as 600 eV for the coarse grid and 800 eV for the finer Fast Fourier Transform (FFT) grid. A Gaussian broadening of 1 eV was used to obtain the density of states (DOS). All simulations shown in this work were carried out using the semilocal Generalized Gradient Approximate (GGA-PBE) exchange-correlation functional \cite{PBE1996}, with our open-source planewave DFT code, \texttt{SHRED} \cite{shred}. A Born-Oppenheimer Molecular Dynamics (BOMD) simulation is performed for each system in an isokinetic-canonical ensemble \cite{Minary03}, in which the nuclei are coupled to a thermostat as they move on an adiabatic potential energy surface. We select ten uncorrelated snapshots from an equilibrated BOMD trajectory and compute the average electronic transport coefficients and estimate the error by the standard deviation over configurations; see Table \ref{tab:tab1}.

\paragraph{Kubo-Greenwood optical conductivity.}
We compute the KG spectra of each system snapshot using its electronic DOS given by the Fermi-Dirac distribution, which, in turn, depends on the density and temperature of the system.
In this work, we introduce a computationally efficient and robust mDFT+KG model. In order to access the electronic transitions between the occupied and unoccupied states in KSDFT+KG, one needs to expand their active space to two to three times the number of ``occupied" orbitals to fully resolve the ground-state electronic density.
This aspect of KSDFT+KG makes it rather memory-intensive and computationally expensive.
Here, states with occupations $\leq 10^{-4}$, $10^{-8}$ are defined as unoccupied for CH and Be, respectively, with a commensurate number of orbitals given in Table \ref{tab:tab1}. This difference in occupation thresholds arises mainly from the number of orbitals required to resolve the tail of the Fermi-Dirac distribution in a computationally tractable manner.

The mDFT+KG method substitutes a large number of deterministic orbitals with fewer stochastic vectors orthogonal to the deterministic subspace \cite{White2020, Sharma2023}.
As shown in Table \ref{tab:tab1} and Figures \ref{fig:ch_kgonsager}(a), \ref{fig:be_kgonsager}(a), \ref{fig:chbe_kgonsager}(a), we partition the occupied density of states such that the upper-lying eigenspectrum is covered stochastically.
Notably, in mDFT+KG, one could still partition the eigenspectrum such that the occupied DOS is spanned almost entirely by deterministic orbitals with a stochastic vector expansion only for unoccupied states. Although this increases the computational cost relative to a more task-optimized orbital partitioning, the approach has the advantage of limiting the active space size and thus avoids the computational scaling of $\mathcal{O}(N^3)$.
The KG formalism in KS-DFT, stochastic DFT, and mDFT is outlined in Section \ref{sec:KG}.


\paragraph{Real-Time TDDFT AC conductivity.}
The finite-temperature ground-state density of a system is instantaneously perturbed by a $\delta-$pulse electric field, homogeneous in space, $E_x = E_0\,\delta(t=0^-)$ and $E_0 = 0.01$ a.u.
A real-time TDDFT--based Ehrenfest dynamics follows in which the electron density evolves according to the time-dependent KS equations, while the nuclei are held fixed, which is a suitable assumption given the targeted simulation time.
A timestep of $0.8$ attoseconds is chosen, and the electron dynamics is performed long enough so that the induced current in the system by the $\delta-$kick field is allowed to equilibrate.
A time series of the electronic current density $\mathbf{J}(t)$ is computed on the fly, multiplied by a Gaussian decay function with a broadening parameter $\Gamma$, and then Fourier transformed to obtain the spectral current density. 
The filtered current signal and the AC conductivity spectra are shown in Supplemental Fig. S1, for different values of $\Gamma$.

\subsection{\label{sec:KG} Kubo-Greenwood formalism in Mixed Stochastic-Deterministic DFT with norm-conserving pseudopotentials}
The Onsager transport coefficients $\mathscr{L}_{mn}$ computed via the KG formula follow from a more universal linear response theory of Kubo \cite{Kubo1957, Greenwood1958, ALLEN2006165, Holst2011}.
The probability electric current ($\bm{\widehat{J}}_{1}$) and heat current density ($\bm{\widehat{J}}_{2}$) operators are defined as \cite{Holst11},
\begin{align}
    \bm{\widehat{J}}_{1} = -\mathsf{i} \widehat{\bm{\nabla}} + \mathsf{i} [{\widehat V_{\text{NL}}, {\widehat {\bm r}}}] ~,\qquad \bm{\widehat{J}}_{2} = \dfrac{\widehat{H}\bm{\widehat{J}}_{1} + \bm{\widehat{J}}_{1} \widehat{H}}{2} - h_e \bm{\widehat{J}}_{1} ~,
    \label{eq:current_operators}
\end{align}
where ${\widehat {\bm r}}$ is the position operator, $h_e$ is the enthalpy per electron, and $\widehat{H}$ is the full norm-conserving Kohn-Sham Hamiltonian of the system, given by
\begin{align}\label{eq:h_ks}
\begin{split}
 \widehat{H} &= -\frac{1}{2}\widehat{\bm{\nabla}}^2 + V(\bm{r}, \bm{r'};\bm{R},\rho_0) ~,\\
 V(\bm{r}, \bm{r'};\bm{R},\rho_0) &= V_{\text{H}}(\bm{r}; \rho_0) + V_{\text{xc}}(\bm{r}; \rho_0) +  V_{\text{ext}}(\bm{r}, \bm{r'}; \bm{R}) ~.
 \end{split}
\end{align}
Here $\rho_0$ is the electron density that minimizes the Kohn-Sham energy,  $V_{\text{H}}$ is the Hartree potential, $V_{\text{xc}}$ is the exchange-correlation potential, and $V_{\text{ext}}$ is the external potential due to electron-ion interactions, with ionic coordinates denoted by $\bm R$. In the pseudopotential approach, this can be split into a local and a nonlocal part, where the latter acts only within the ``core" regions of the atomic sites,
\begin{align}
    V_{\text{ext}}(\bm{r}, \bm{r'}; \bm{R}) = V_{\text{ext},\text{loc}}(\bm{r}; \bm{R}) + V_{\text{ext},\text{NL}}(\bm{r}, {\bm r'}; \bm{R}) ~,
\end{align}
where $V_{\text{ext},\text{NL}}$ is the nonlocal potential due to the approximate treatment of the ion-electron interaction.

In the KG (single-particle Kohn-Sham) approach, the Onsager coefficients in atomic units take the form,
\begin{align}
    \mathscr{L}_{mn}(\omega) &= (-1)^{m+n} \dfrac{2\pi}{3\,\omega\,\Omega} \sum_{i,j}\left(\dfrac{\epsilon_i + \epsilon_j}{2} - h_e \right)^{m+n-2} \nonumber\\
    & \quad \times \big|\langle\psi_i|\bm{\widehat{J}}_{1}| \psi_j\rangle\big|^2 \times [f(\epsilon_i) - f(\epsilon_j)] \nonumber\\
    & \quad \times \delta(\epsilon_j-\epsilon_i-\omega) ~,
    \label{eq:onsager}
\end{align}
for electronic transitions between deterministic eigenstates $\ket{\psi_i}$ with eigenenergies $\epsilon_i$, where $\Omega$ is the cell volume, $f(\epsilon)$ is the Fermi-Dirac distribution function, and the Dirac delta distribution is understood to be approximated by a Lorentzian or a Gaussian. The electrical and thermal conductivity tensors follow from the Onsager coefficients:
\begin{align}
    \sigma (\omega) = \mathscr{L}_{11} ~, \quad \kappa (\omega) = \dfrac{1}{T}\left(\mathscr{L}_{22} - \dfrac{\mathscr{L}_{12}\mathscr{L}_{21}}{\mathscr{L}_{11}}\right) ~.
\end{align}
Numerically, $\mathscr{L}_{mn}$ can be computed as a Fourier transform of the two-point time-domain current-current correlation function with a finite broadening parameter $\Gamma$, given as
\begin{align}
    \mathscr{L}_{mn}(\omega) &= (-1)^{m+n} \dfrac{2\pi}{3\,\omega\,\Omega} \textsf{Re} \int_{0}^{\infty} dt\, e^{\mathsf{i}(\omega +\mathsf{i}\Gamma(t))t} C_{mn}(t) ~, \label{eq:onsager_current}
\end{align}
where,
\begin{align}
    C_{mn}(t)\mid_{\textbf{det.}} &= \sum_{i,j} \left( \dfrac{\epsilon_i + \epsilon_j}{2} - h_e\right)^{m+n-2} \nonumber\\ & \times |\bra{\psi_i}\bm{\widehat{J}}_1  \ket{\psi_j}|^{2} 
      \times [f(\epsilon_i)-f(\epsilon_j)]e^{-\mathsf{i}(\epsilon_j-\epsilon_i)t} ~.
    \label{eq:onsager_det}
\end{align}
For a Lorentz (resp. Gaussian) broadening, $\Gamma(t) = \gamma_0$ (resp. $\frac{1}{4}\gamma_0^2 t$), such that $\mathsf{Re}(\Gamma) > 0$. In Eq. \eqref{eq:onsager_det}, the subscript \textbf{det.} indicates that the current-current correlation function is resolved using deterministic Kohn-Sham eigenstates. In fact, we recall the more general definition of the current-current correlation function as,
\begin{align}
    \label{eq:C_general}
    C_{mn}(t) &= 2 \,\textsf{Im\,}[\textsf{Tr}\{\bm{\widehat{J}}_m(t) \cdot \bm{\widehat{J}}_n f(\widehat{H})\}] ~, \\\nonumber
    \quad \bm{\widehat{J}}_m(t) &= e^{\mathsf{i}\widehat{H}t}{\bm{\widehat{J}}}_m e^{-\mathsf{i}\widehat{H}t} ~. \nonumber
\end{align}
Here, $\bm{\widehat{J}}_m(t)$ is the time-dependent current operator in the Heisenberg picture, and $f(\widehat{H})$ is the finite-temperature Kohn-Sham-Mermin single-particle density matrix \cite{Mermin1965}. Using Hutchinson's stochastic trace formula \cite{Hutchinson1990}, we can replace the trace operation with an expectation over randomly generated stochastic vectors, $\langle \bm r \ket{{\chi_{\alpha}}} \equiv \big(\frac{1}{\Delta N_{\chi}}\big)^{1/2} e^{\mathsf{i}2\pi\theta_{\alpha}({\bm r})}$, where $\theta_{\alpha}({\bm r})$ is a uniformly distributed random number between 0 and 1, $N_{\chi}$ is the number of stochastic vectors, and $\Delta$ denotes the grid volume. This yields a stochastic-trace definition of the current-current correlation function,
\begin{align}
    \left.C_{mn}(t)\right|_{\textbf{sto.}} &= 2 \,\textsf{Im\,} \sum_{\alpha} \bra{{\chi}_\alpha} f^{\frac{1}{2}}(\widehat{H}) \bm{\widehat{J}}_m(t) \cdot \underbrace{\bm{\widehat{J}}_n f^{\frac{1}{2}}(\widehat{H})\ket{{\chi}_\alpha}}_{\equiv \ket{\bm{J}^{\chi}_{n,\alpha}}} ~,
    \label{eq:current_sto}
\end{align}
where symmetric multiplication of stochastic vectors via the matrix function $f^{\frac{1}{2}}(\widehat{H}) \ket{\chi_{\alpha}} \equiv \ket{\mathcal{X}_{\alpha}}$ is numerically preferred and was previously described in \cite{Cytter2019}. 
Since these filtered stochastic vectors are not eigenvectors of the Kohn-Sham Hamiltonian, the additional multiplication of $\bm{\widehat{J}}_n f^{\frac{1}{2}}(\widehat{H})\ket{{\chi_{\alpha}}} \equiv \ket{{\bm {J}^\chi_{n,\alpha}}} $ and $\ket{\mathcal{X}_{\alpha}}$ by the propagator, $e^{\mathsf{-i}{\widehat H}t}$, must be carried out numerically \cite{Cytter2019}. 
Thus, for every stochastic vector, up to seven vectors, $ \ket{ {J^\chi_{n\in \{1,2\}, \{x,y,z\}, \alpha}}}$ and $\ket{\mathcal{X}_\alpha}$, are propagated.
To solidify the connection between the approaches, we note that Eq. \eqref{eq:C_general} is general to either a stochastic or deterministic trace by replacing sum over $\ket{{\chi}}$'s with a converged number of $\ket{{\psi}}$'s in Eq \eqref{eq:current_sto}, i.e., using deterministic orbitals to construct resolution of identity. To arrive at Eq. \eqref{eq:onsager_det} an additional resolution of identity is introduced ($e^{-\mathsf{i}{\widehat{H}t}} \equiv \sum_j e^{-\mathsf{i}{{\epsilon_j}t}}\ket{\psi_j}\bra{\psi_j}$) which removes explicit propagation of $\ket{{\bm J}^\psi}$, but requires computation of both occupied and unoccupied orbitals. 

In mDFT, we utilize both deterministic KS and stochastic orbitals using the mixed trace / mixed resolution of identity:
\begin{align}
    \label{eq:mixed_ROI}
    \widehat{I} &\approx \sum_{i=1}^{N_\psi} \ket{\psi_i}\bra{\psi_i} + \sum_{\alpha=1}^{N_\chi} \ket{{\chi}_\alpha'}\bra{{\chi}_\alpha'} ~,
\end{align}
where \cite{White2020},
\begin{align}
\ket{{\chi}'_\alpha} := \ket{{\chi}_\alpha} - \sum_{i=1}^{N_\psi} \ket{\psi_i}\bra{\psi_i} {\chi}_\alpha \rangle ~.
\end{align}
 It is immediately obvious that utilization of Eq. \eqref{eq:mixed_ROI} to take the trace in Eq. \eqref{eq:C_general}, yields deterministic and stochastic contributions to the current-current correlations. To avoid propagation of $\ket{{\bm {J}^\psi}}$ \textit{in the deterministic trace}, we again insert the mixed resolution of identity, which returns Eq. \eqref{eq:onsager_det} with fewer deterministic orbitals and an additional cross-term arising from mixed transitions between the deterministic eigenstates and stochastic orbitals:
\begin{align}
    \label{eq:current_mixed}
    C_{mn}(t)\mid_{\textbf{mix.}} &= \sum_{i} \sum_{\alpha} \bra{\psi_i}\bm{\widehat{J}}_m e^{-\mathsf{i}\widehat{H}t} \ket{{\chi}'_{\alpha}} \\\nonumber & \quad \quad \quad \quad \quad \quad \times  \bra{{\chi}'_{\alpha}} \bm{\widehat{J}}_n \ket{\psi_i} f(\epsilon_i)e^{\mathsf{i}\epsilon_it}  ~, 
\end{align}
\begin{align}
    \label{eq:C-tot}
    C_{mn}(t) = C_{mn}(t)\mid_{\textbf{det.}} + C_{mn}(t)\mid^{\perp}_{\textbf{stoc.}} + C_{mn}(t)\mid_{\textbf{mix.}}~.
\end{align}
Utilizing the same set of stochastic vectors as for Eq. \eqref{eq:current_sto}, this only requires a computation of matrix elements between deterministic and stochastic vectors for initial and time-evolved unfiltered stochastic vectors.
We employ the unitary Short Iterative Lanczos method to propagate the orbitals in time \cite{Park1986}. Alternatively, the time-evolution operator could also be evaluated via Chebyshev expansions \cite{Kosloff1988, Cytter2019}.


\subsection{\label{sec:tddft}Mixed Stochastic-Deterministic Time-Dependent Density Functional Theory (real-time TDDFT)}
The optical AC conductivity ($\sigma(\omega)$) can also be calculated as a dynamical response of the system to a direct perturbation. An electric field pulse $\bm{E} = E_x(t) \mathbf{\widehat{x}} = E_0 \delta(t - t_{0})\mathbf{\widehat{x}}$ (with $E_0 \approx  0.01$ a.u.), homogeneous in space, is applied at $t=t_{0} = 0^-$, resulting in a current density $J_y$ in the system, given by
\begin{align}
    \sigma_{yx}(\omega) &= \dfrac{\partial J_y (\omega)}{\partial E_{x}} ~, \quad J_y(\omega) = \int_0^\infty dt\, e^{\mathsf{i}\omega t} J_y(t) \, e^{-\frac{\gamma_0^2 t^2}{4}} ~.
    \label{eq:opticalctonductivity}
\end{align}
In this approach, the electronic wavefunctions follow the (adiabatic) real-time time-dependent Kohn-Sham equations (rt-TDDFT) in the velocity gauge given by
\begin{align}
    \label{eq:tdks}
    \mathsf{i}\dfrac{\partial \varphi(\bm{r}, t)}{\partial t} &= \widehat{H}(t) \varphi(\bm{r}, t) ~, \\
    \widehat{H}(t) &\equiv \dfrac{1}{2}\big(-\mathsf{i}\widehat{\bm \nabla} + {\bm A(t)}\big)^2 + V(\bm{r}, \bm{r'}; \bm{R}, \rho(t)) ~,\nonumber
\end{align}
where $\varphi$ can be initialized as either a deterministic Kohn-Sham eigenvector, $\psi$, or filtered complementary stochastic vector, $\mathcal{X}$ \cite{White22}.  The electronic current calculated by mixed TDDFT is then given by:
\begin{align}\begin{split}
    {\bm J(t)} &= \sum_{i=1}^{N_\psi} f(\epsilon_i)\bra{\psi_i(t)}-\mathsf{i}\widehat{\bm \nabla} + {\bm A(t)} \ket{\psi_i(t)} \\ 
    &\quad + \sum_{\alpha=1}^{N_\chi} \bra{\mathcal{X}_\alpha(t)}-\mathsf{i}\widehat{\bm \nabla} + {\bm A(t)} \ket{\mathcal{X}_\alpha(t)} ~.
    \end{split}
\end{align}
%
Here, $E_x(t) = E_{0}\delta(t-t_0)$. We choose a gauge in which all components of $\bm{A}(t)$ except $A_x(t)$ vanish. This implies that $A_x(t) = -E_{0}\Theta(t-t_0)$.The direct perturbation approach is versatile having been applied to the calculation of not only electrical conductivity, but also dynamic structure factors, and nonadiabatic Born effective charges \cite{Baczewski16, Kononov2022, Andrade2018, Ramakrishna23, Ramakrishna24, White22, White2024, White25, Sharma2025}.



In this work, we have a two-pronged focus, first on the use of mixed DFT--based KG in regimes where KSDFT+KG approach would be rendered inadequate due to a finite number of KS orbitals afforded by the simulation, and second is to make connections to optical response properties via rt-TDDFT dynamics based on the time-evolution of strictly occupied orbitals. The latter offers the natural advantage of not requiring any extra ``unoccupied" orbitals or propagation of $ \ket{\bm J_{\chi}}$ vectors, but does require propagation of the occupied deterministic orbitals. Note that in Eq. \eqref{eq:tdks}, it is the time-dependence of the electron density in the Hartree and exchange-correlation potentials, which provides the difference between the TDDFT and KG levels of theory within the linear-response regime.  


\subsection{\label{sec:paw-tddft} mDFT, TD-mDFT and Kubo-Greenwood in the PAW framework}
The PAW approach provides a route towards the efficient treatment of sharp features in the Kohn-Sham orbitals when using real or planewave basis sets by per-computation of PAW datasets, consisting of on-site atomic orbitals $\vert \phi_i \rangle$ and atom-centered projector function $\vert P \rangle$ which form a dual-basis with smoothed on-site atomic orbitals, $\vert {\widetilde \phi}_i \rangle$ ($\langle \widetilde {\phi}_i \vert P_j \rangle = \delta_{ij}$). Here, we review only the practical alterations of the KG and real-time TD-DFT methods. For a detailed discussion on the PAW method, see Refs. \cite{Blochl94, Rostgaard09,Kresse99} and the supplemental of Ref. \cite{Sharma2023}. 

Application of the PAW formalism leads to the transformed (pseudized) Hamiltonian and Identity operator:
\begin{align}
{\widehat {\widetilde H}} &= \widehat{H} +  \sum_a \sum_{n,m} \vert P^a_n \rangle {D}_{nm} \langle P^a_m \vert ~,\\
\widehat{\!\widetilde I} &\equiv {\widehat S} =  \widehat{I} +  \sum_a \sum_{n,m} \vert P^a_n \rangle {s}_{nm} \langle P^a_m \vert ~, \nonumber\\
{D}_{nm} &= \langle \phi_n \vert {\widehat H} \vert \phi_m \rangle - \langle {\widetilde \phi}_n  \vert {\widehat H} \vert {\widetilde \phi}_m \rangle ~,\nonumber\\
   {s}_{nm} &= \langle \phi_n \vert \phi_m \rangle - \langle {\widetilde \phi}_n   \vert {\widetilde \phi}_m \rangle ~, \nonumber
\end{align}
where $n,m$ index the atomic orbital and $a$ indexes the atom center. Furthermore, the approximate mixed resolution of identity as shown in Eq. \eqref{eq:mixed_ROI}, is now replaced by the mixed resolution of the inverse overlap operator, ${\widehat S}^{-1}$. The deterministic orbitals are solutions to a generalized eigenvalue problem, and stochastic vectors must be pre-multiplied by the root-inverse of PAW overlap \cite{Li2020, Sharma2023}:
\begin{align}
\label{eq:mixed_ROI_PAW}
\widehat{S}^{-1} &\approx \sum_{i=1}^{N_\psi} \ket{\widetilde{\psi}_i}\bra{{\widetilde \psi}_i} + \sum_{j=1}^{N_\chi} \ket{\widetilde{\chi}'_j}\bra{\widetilde{\chi}'_j} ~,  \\
{\widehat {\widetilde H}} \vert {\widetilde \psi}_i \rangle & = {\widehat S} \vert {\widetilde \psi}_i \rangle \epsilon_i ~, \quad \ket{\widetilde{\chi}'_j}  \equiv \widehat{S}^{-{1/2}} \ket{{\chi}'_j} ~.\nonumber 
\end{align}

For calculations of current and heat current density, the transformed operators read as follows. 
\begin{align}\begin{split}
    \bm{\widehat{\!\!\widetilde J}}_{1} = -\mathsf{i}\bm{\widehat \nabla} + 
    \sum_a \sum_{i,j} \vert P^a_i \rangle {\bm T}_{ij} \langle P^a_j \vert  ~,\\
    \bm{\widehat{\!\!\widetilde J}}_{2} = \dfrac{\widehat{\widetilde H}\, \widehat{S}^{-1}\,\bm{\widehat{\!\!\widetilde J}}_{1} + \, \,\bm{\widehat{\!\!\widetilde J}}_{1} \widehat{S}^{-1} \widehat{\widetilde H}}{2} - h_e \,\,\bm{\widehat{\!\!\widetilde J}}_{1} ~,  \\
    {\bm T}_{ij} = \langle \phi_i \vert -\mathsf{i}\bm{\widehat \nabla} \vert \phi_j \rangle - \langle {\widetilde \phi}_i  \vert -\mathsf{i}\bm{\widehat \nabla} \vert {\widetilde \phi}_j \rangle ~,
\end{split}
    \label{eq:paw_current_operators}
\end{align}
where ${\bm T}$ is the on-site momentum operator. Transformed products of operators can be replaced by the product of transformed operators separated by a $\widehat S^{-1}$, i.e.,
\begin{align}
{\widehat {\widetilde{AB}}} = {\widehat{\!\widetilde A}}{\widehat S}^{-1}{\widehat{\widetilde B}} ~.
\end{align}
Matrix functions, e.g., the density matrix and propagators appearing in Eqs. \eqref{eq:C_general}-\eqref{eq:current_mixed}, are transformed as 
\begin{align}
\widetilde{f(\widehat{H})} \mapsto f({\widehat S}^{-1} \widehat{\widetilde{H}}) ~.
\end{align}
Note that Eq. \eqref{eq:mixed_ROI_PAW} is applied when a low-rank form of $\widehat{S}^{-1}$ is required, e.g., analogous to using approximate resolution of identity in the norm-conserving case. In other cases, $\widehat S^{-1}$ multiplies a vector, which is the solution to a linear system of equations involving ${\widehat S}$, or using the Woodbury formula \cite{Woodbury:1949,Woodbury:1950,Higham:2002,Golub2013-dz} for the inversion of a diagonal plus low-rank matrix. 


\section{\label{sec:results}Results}
\begin{figure*}[t!]
    \includegraphics[width=6.5 in]{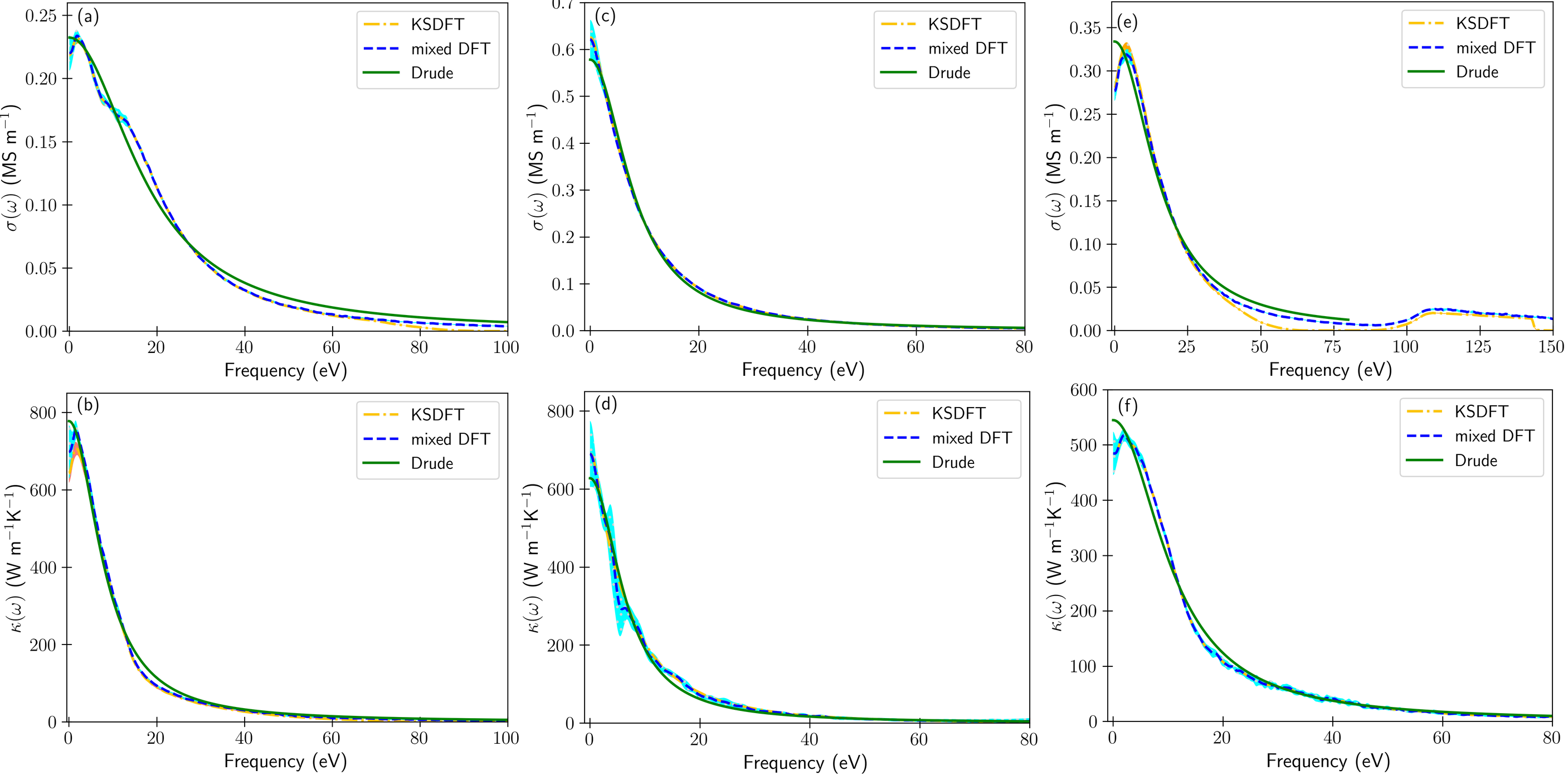}
    \caption{Optical conductivity spectra as a function of the photon energy obtained using the Kubo-Greenwood formalism within KS-DFT (yellow dashdotted line) and mDFT (blue dashed line). Electrical and Thermal Conductivity (a, b): CH mixture at $(\rho, T)= (0.9 \text{ g/cm}^3, 7.8 \text{ eV})$, (c, d): Be at $(\rho, T)= (1.84 \text{ g/cm}^3, 4.4 \text{ eV})$ and, (e, f): CH/Be ternary mixture at $(\rho, T) = (1.37 \text{g/cm}^3, 5.0 \text{ eV})$, respectively. A Drude-model fit (green solid line) to the mDFT conductivity is included as a reference.}
    \label{fig:chbe_kg}
\end{figure*}

\begin{figure*}[t!]
    \includegraphics[width=5 in]{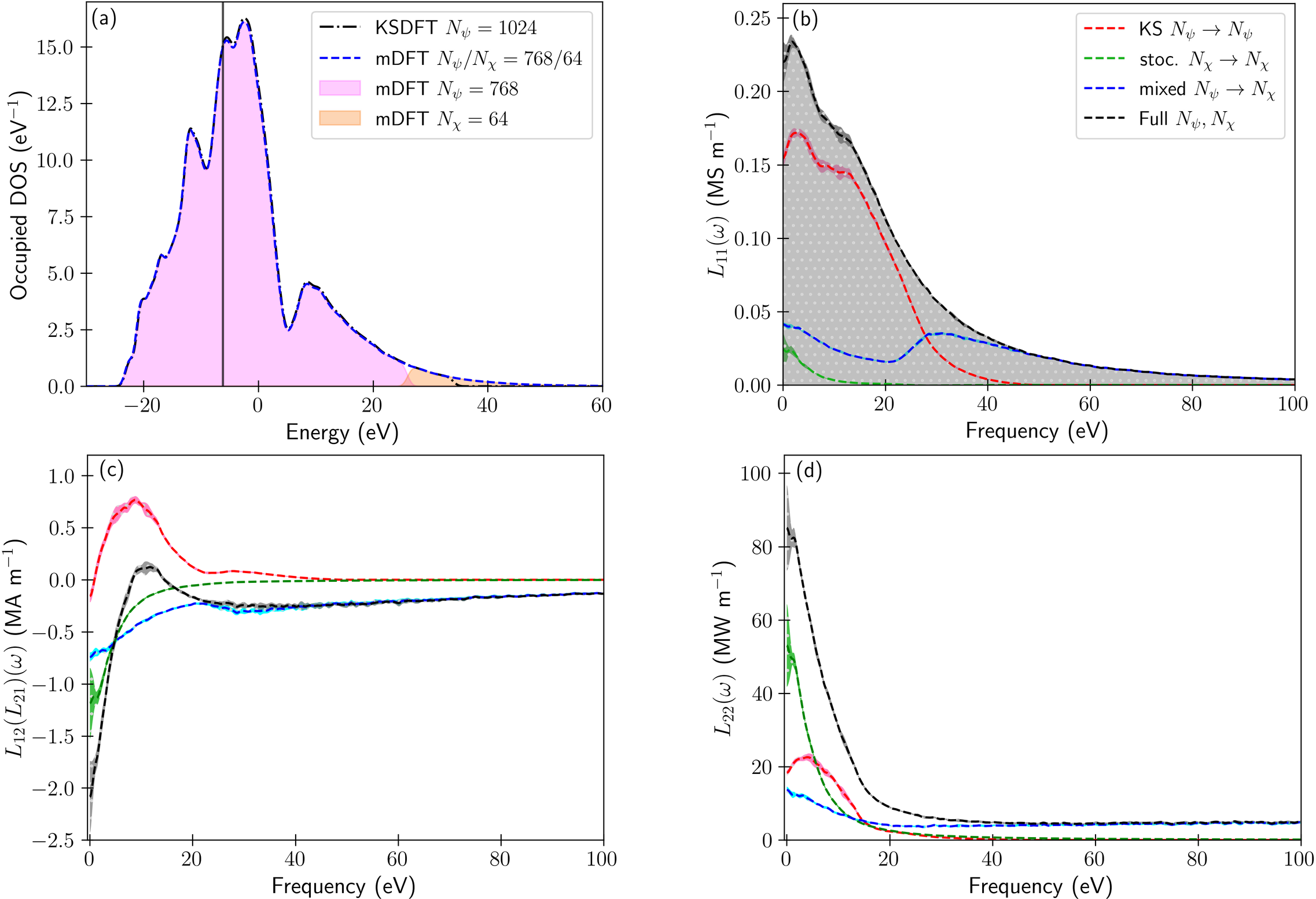}
    \caption{CH mixture at $(\rho, T)= (0.9 \text{ g/cm}^3, 7.8 \text{ eV})$: (a) Occupied density of states (DOS) obtained with KS-DFT and mDFT. The pink and orange-shaded regions indicate the contributions of deterministic KS and stochastic orbitals, respectively, to the mixed DOS. (b)-(d) Frequency-dependent Onsager coefficients $L_{mn}(\omega)$ decomposed into transitions among Kohn-Sham ($N_\psi$), stochastic ($N_\chi$), and mixed deterministic-stochastic ($N_\psi, N_\chi$) orbitals. The full $L_{mn}(\omega)$ shown in black dashed lines comprises transitions amongst all orbitals.} 
    \label{fig:ch_kgonsager}
\end{figure*}

\begin{figure*}[t!]
    \includegraphics[width=5 in]{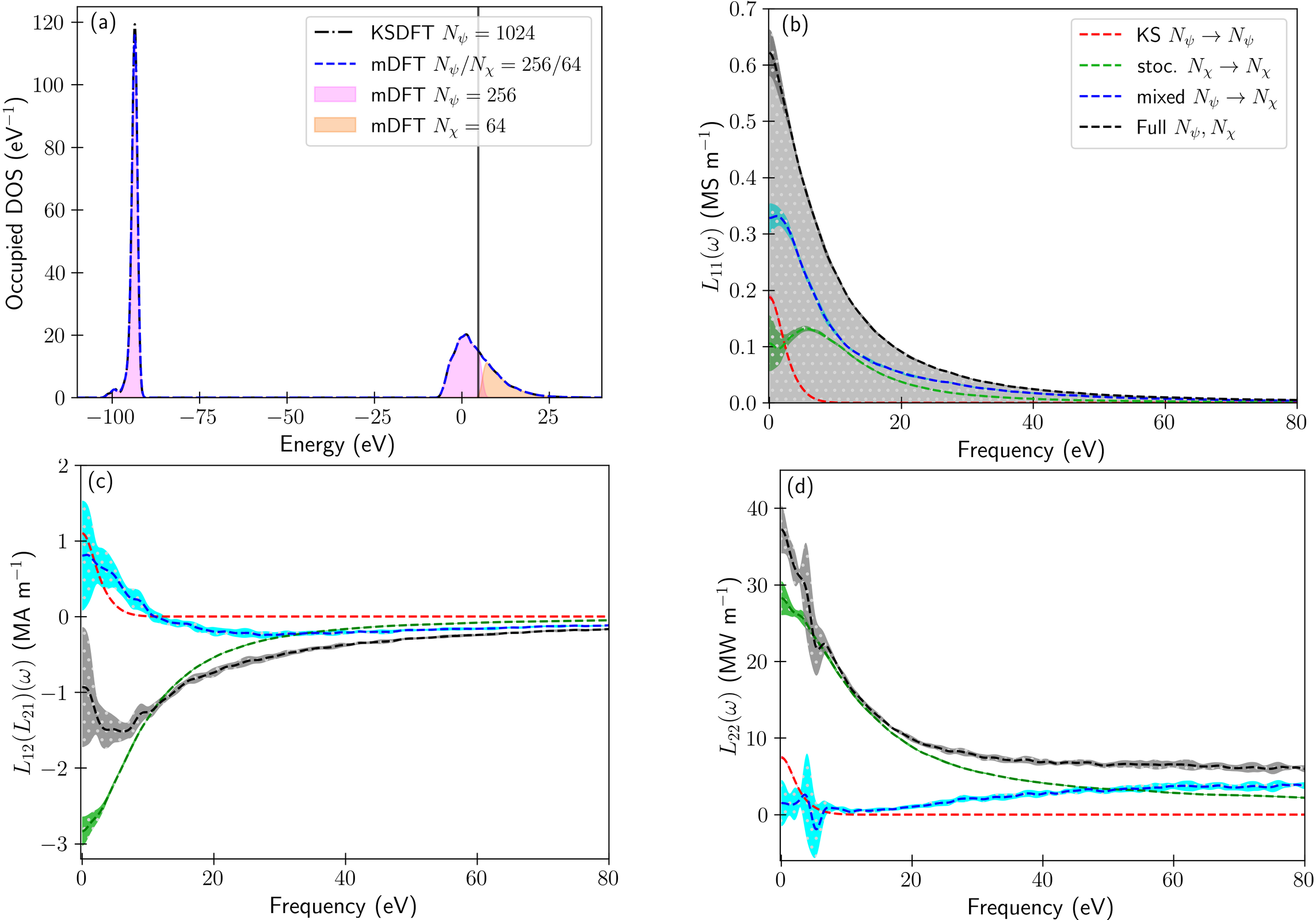}
    \caption{Be at $(\rho, T)= (1.84 \text{ g/cm}^3, 4.4 \text{ eV})$: (a) Occupied density of states (DOS) obtained with KS-DFT and mDFT. The pink and orange-shaded regions indicate the contributions of deterministic KS and stochastic orbitals respectively to the mixed DOS. (b)-(d) Frequency-dependent Onsager coefficients $L_{mn}(\omega)$ decomposed into transitions among Kohn-Sham ($N_\psi$), stochastic ($N_\chi$), and mixed deterministic-stochastic ($N_\psi, N_\chi$) orbitals. The full $L_{mn}(\omega)$ shown in black dashed lines comprises transitions amongst all orbitals.} 
    \label{fig:be_kgonsager}
\end{figure*}

\begin{figure*}[t!]
    \includegraphics[width=5 in]{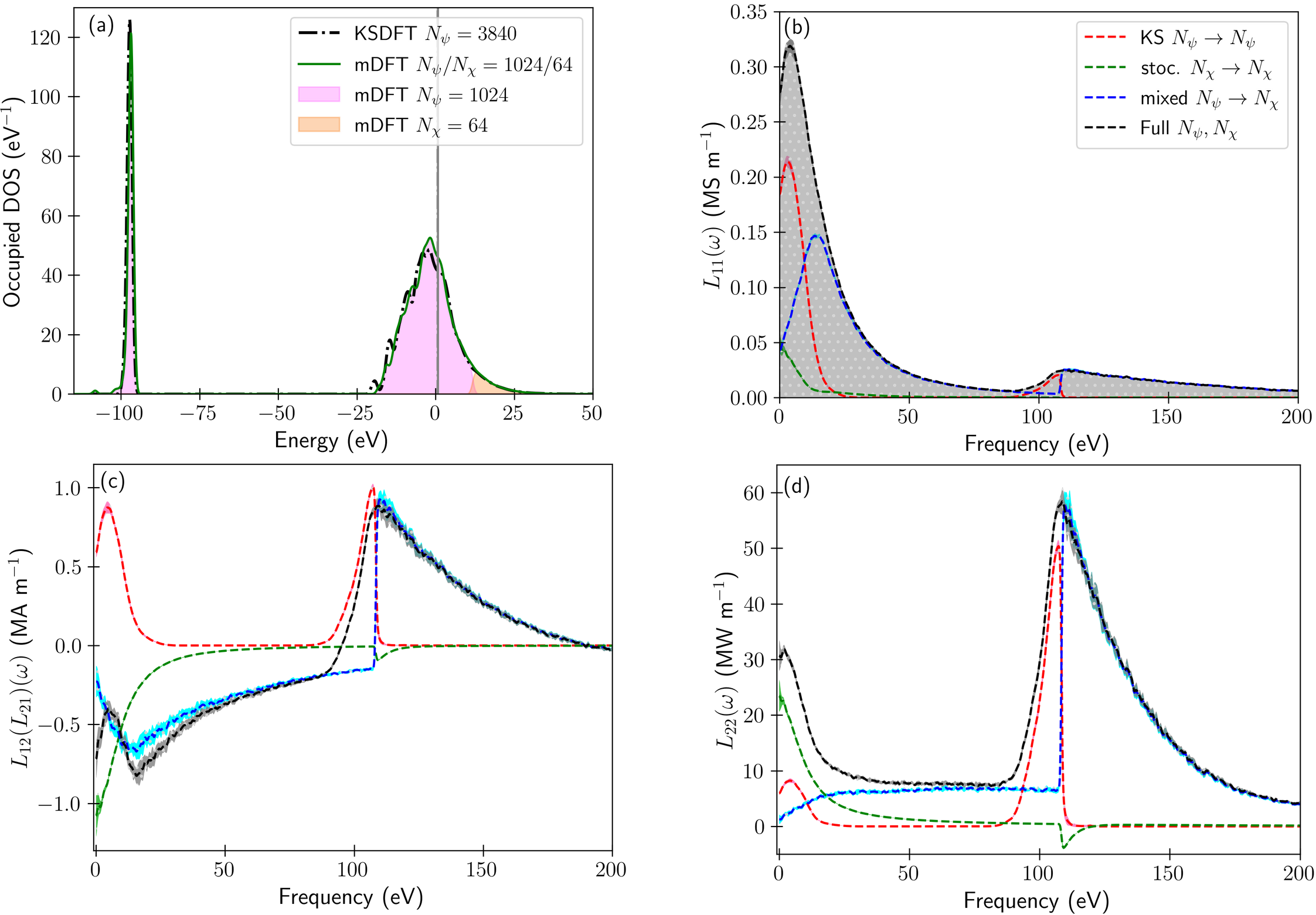}
    \caption{CH/Be at $(\rho, T)= (1.37 \text{ g/cm}^3, 5.0 \text{ eV})$: (a) Occupied density of states (DOS) obtained with KS-DFT and mDFT. The pink and orange-shaded regions indicate the contributions of deterministic KS and stochastic orbitals respectively to the mixed DOS. (b)-(d) Frequency-dependent Onsager coefficients $L_{mn}(\omega)$ decomposed into transitions among Kohn-Sham ($N_\psi$), stochastic ($N_\chi$), and mixed deterministic-stochastic ($N_\psi, N_\chi$) orbitals. The full $L_{mn}(\omega)$ shown in black dashed lines comprises transitions amongst all orbitals.} 
    \label{fig:chbe_kgonsager}
\end{figure*}

In mDFT, a typical partitioning of the eigenspectrum into maximally occupied low-energy deterministic ($N_\psi$) and higher-energy stochastic ($N_\chi$) segments is shown by pink and orange-shaded regions in Figures \ref{fig:ch_kgonsager}, \ref{fig:be_kgonsager}, \ref{fig:chbe_kgonsager} (a). The occupied DOS is compared for KS-DFT and mDFT and it is rightly found that large $N_\psi$ orbitals are required to cover the DOS deterministically, while one can considerably tune down $N_\psi$ in favor of $N_\chi$ in mDFT to obtain a converged spectrum of states. In fact, the agreement between KS-DFT and mDFT is qualitatively represented for a variety of systems in Figures \ref{fig:ch_kgonsager}, \ref{fig:be_kgonsager}, \ref{fig:chbe_kgonsager} (a) alongside quantifiable measures listed in Table \ref{tab:tab1}. Evidently, we find a very good agreement between the DC electrical and thermal conductivities obtained from the two methods.
For a more detailed comparison of observables, such as energies, pressures, and forces, in systems relevant to warm dense matter, we point the reader to \cite{Sharma2023, White2020}.

To determine the optical response properties of a symmetric carbon-hydrogen mixture under physical conditions similar to the plastic foam found in fuel capsules \cite{ICF2020, ICF2021} as shown in Figure \ref{fig:ch_kgonsager}, we find that it requires at least $N_\psi=2560$ bands with KSDFT+KG, see Table \ref{tab:tab1}.
However, introducing stochastic vectors ($N_\chi=64$) with mixed DFT+KG, we obtain converged electrical and thermal conductivity within a suitable error of uncertainty at reduced $N_\psi=768$ bands and hence lower computational costs.
The KS-DFT- and mDFT- derived electrical and thermal conductivity spectra for CH are compared in Figure \ref{fig:chbe_kg}(a), (b), respectively, where the cyan-shaded region represents the standard deviation in mixed DFT due to different stochastic vectors chosen over ten different snapshots. The maximum deviation occurs in the low-frequency region and increases as $\omega \to 0$.
In each of the cases, the conductivity spectra are extrapolated to the static limit $\omega\to0$, listed in Table \ref{tab:tab1}.
Since all systems studied in this work exhibit a metallic character at the given $(\rho, T)$, a Drude model fit of the conductivity is also shown as a solid green line in Figure \ref{fig:chbe_kg}.
In the mixed DFT+KG formalism, we explicitly compute the Onsager coefficients $\mathcal{L}_{mn}$ leading to the conductivity as described in Section \ref{sec:KG}. This implies that one can decompose each charge transport component $\mathcal{L}_{mn}$ into contributions from transitions among different types of orbitals. Figures \ref{fig:ch_kgonsager}(b)-(d) illustrate the electronic transitions associated with KS $\rightarrow$ KS, stochastic $\rightarrow$ stochastic, and mixed KS-stochastic orbitals, and with their uncertainties indicated by the shaded regions. All the orbital contributions taken together produce the full Onsager coefficient. This representation allows us to compare the extent to which a set of stochastic orbitals contributes to a region of the optical spectrum and its DC ($\omega \to 0$) limit. Of course, the $\mathcal{L}_{11}$ decomposition in Figure \ref{fig:ch_kgonsager}(b) directly yields the orbital contributions to the AC electrical conductivity $\sigma(\omega)$.

Another system of interest is warm-dense beryllium at its preferred solid-state density of 1.84 g/cm$^3$ and a temperature of 4.4 eV. The PAW potential for Be includes its core $1s^2$ electrons resulting in an inner low-lying peak visible in the occupied DOS, see Figure \ref{fig:be_kgonsager} (a). 
We carry out a similar orbital-type decomposition of the Onsager coefficients in Figure \ref{fig:be_kgonsager}(b)–(d), and observe a higher contribution from stochastic and mixed orbital transitions compared to the previous example of CH mixture. This behavior is highly expected, given the larger ratio of stochastic to KS orbitals employed in the mixed DFT+KG treatment of Be. The corresponding conductivity spectra appear in Figure \ref{fig:chbe_kg}(c), (d), showing an excellent agreement between the KS and mixed DFT methods up to a suitable uncertainty, also evidenced by the static DC values in Table \ref{tab:tab1}.

Next, we explore an instructive example of CH/Be mixture at $T=5$ eV obtained by averaging the densities of the two systems studied so far, at 1.37 g/cm$^3$.
An individual chemical componentwise split of the charge transport properties of CH/Be appeared recently in our work focusing on effective charge determination for disordered metals \cite{Sharma2025}.
The occupied DOS shown in Figure \ref{fig:chbe_kgonsager}(a) has characteristics of CH and Be as captured by the broader peak around the chemical potential and the deeper sharp peak located at around $-$100 eV. The decomposition of Onsager coefficients shows a long shoulder peak attributed to the transitions in Be, in Figure \ref{fig:chbe_kgonsager}(b)-(d). While the low-frequency spectra are dominated by transitions among KS orbitals, in contrast, the higher-frequency region of the spectra mostly comprises stochastic and mixed transitions. It is worth noting that Figure \ref{fig:chbe_kg}(e) shows a cutoff in the high-frequency limit of the spectra with KS-DFT due to the finite number of total (occupied + unoccupied) deterministic orbitals allowed on account of computational feasibility. However, in mixed DFT, stochastic vectors provide a smoother tail at higher frequencies while preserving the low-frequency accuracy of KS-DFT.

\begin{figure}
    \includegraphics[width=3 in]{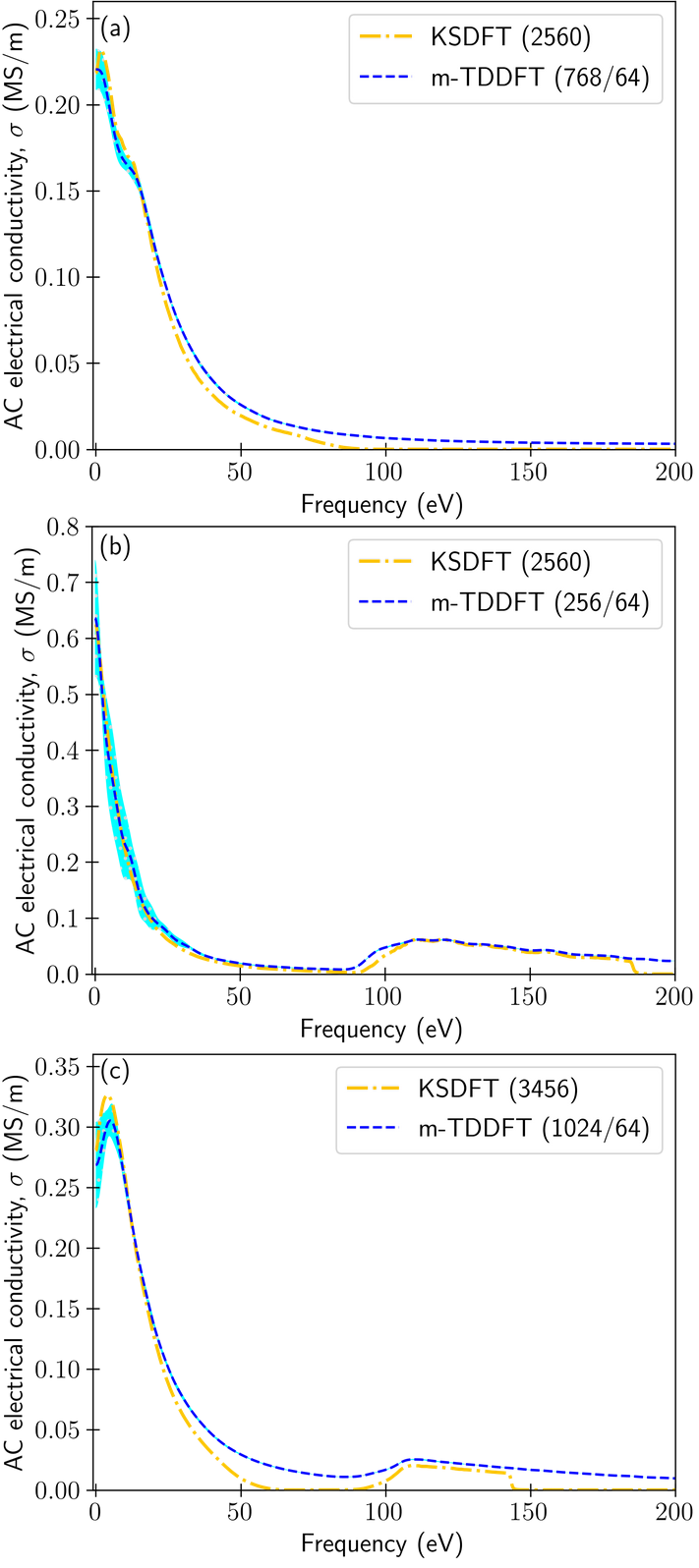}
    \caption{Comparison of electrical AC conductivity spectra obtained from Kubo-Greenwood + Kohn-Sham DFT and real-time Time-Dependent mixed DFT--based Ehrenfest dynamics for: (a) CH $(1:1)$ mixture, (b) Be and, (c) CH/Be ternary mixture. The cyan-shaded region shows the uncertainty from mixed DFT.}
    \label{fig:tddft_conductivity}
\end{figure}
As discussed in Section \ref{sec:tddft}, we also derive optical (electrical) conductivity as a response of the system to a perturbation such as a macroscopic electric field. In Figure \ref{fig:tddft_conductivity}, we compare conductivity spectra obtained from the time evolution of occupied mixed deterministic-stochastic orbitals following a $\delta-$ field pulse vs. Kubo-Greenwood with a large number of (occupied + unoccupied) KS orbitals for the three warm-dense systems studied in this work.
The spectra indicate a good agreement between the two methods within the uncertainty bounds (cf. cyan-shaded region in Figure \ref{fig:tddft_conductivity}) calculated as a standard deviation over ten equilibrated snapshots and hence comprising error of stochasticity. The corresponding static DC values reported in Table \ref{tab:tab1} show a good agreement with the $\sigma(\omega\to0)$ values of KSDFT+KG and mixed DFT+KG.
While the results for CH/Be ternary mixture in Figure \ref{fig:tddft_conductivity}(c) include the conductivity response of all the elements, in Reference \cite{Sharma2025}, we presented the first-ever decomposition of a macroscopic system property into its chemical constituents.
An example of an electronic current density decay in the time domain during rt-TDDFT dynamics appears for CH/Be in the Supplementary Information, Figure S1; such a time signal is Fourier transformed to yield the conductivity response spectra shown in Figure \ref{fig:tddft_conductivity}.
Although the rt-TDDFT dynamics directly provides access to electrical conductivity, there is currently no straightforward prescription for obtaining thermal conductivity from rt-TDDFT.




\section{\label{sec:conclusion}Conclusion}
Over the past few decades, DFT-derived methods have established their usefulness for describing properties of warm dense matter and hot dense plasmas in the most physically `first-principles' fashion, accessible via modern computational architectures for these rather challenging regimes \cite{Ligneres2005, Carter2008, HuangCarter2010, Chen2013, BaerRabani2013, Karasiev2014, Sjostrom2015, Chen2015, Gao2015, Ding2018, White2018, Cytter2018, Fabian2019, White2020, Jiang2022, Pavanello2022, PavanelloTrickey2023, 2025roadmapwarmdensematter}. 
However, many of these methods were not explicitly designed to capture electron transport, which requires some knowledge of the band structure (orbital) properties of the system. 
In this work, we demonstrate the utilization of mixed stochastic-deterministic DFT-based charge-transport models for extreme states of matter. To further verify and validate the use of mixed deterministic (KS) and stochastic orbitals, we evaluate Onsager coefficients from intra- and inter- band transitions between the two orthogonal sets of orbitals. Taken together, these contributions reconcile the full transport coefficient and hence, the conductivity over high and low frequencies.

\begin{acknowledgments}
This work was supported by the U.S. Department of Energy through the Los Alamos National Laboratory (LANL). Research presented in this article was supported by the Laboratory Directed Research and Development program, projects number 20230322ER and 20230323ER, of LANL. We acknowledge the support of the Center for Nonlinear Studies (CNLS). This research used computing resources provided by the LANL Institutional Computing. Los Alamos National Laboratory is operated by Triad National Security, LLC, for the National Nuclear Security Administration of U.S. Department of Energy (Contract No. 89233218CNA000001).
\end{acknowledgments}

\bibliographystyle{apsrev4-2}
\bibliography{references}

\begin{thebibliography}{78}%
\makeatletter
\providecommand \@ifxundefined [1]{%
 \@ifx{#1\undefined}
}%
\providecommand \@ifnum [1]{%
 \ifnum #1\expandafter \@firstoftwo
 \else \expandafter \@secondoftwo
 \fi
}%
\providecommand \@ifx [1]{%
 \ifx #1\expandafter \@firstoftwo
 \else \expandafter \@secondoftwo
 \fi
}%
\providecommand \natexlab [1]{#1}%
\providecommand \enquote  [1]{``#1''}%
\providecommand \bibnamefont  [1]{#1}%
\providecommand \bibfnamefont [1]{#1}%
\providecommand \citenamefont [1]{#1}%
\providecommand \href@noop [0]{\@secondoftwo}%
\providecommand \href [0]{\begingroup \@sanitize@url \@href}%
\providecommand \@href[1]{\@@startlink{#1}\@@href}%
\providecommand \@@href[1]{\endgroup#1\@@endlink}%
\providecommand \@sanitize@url [0]{\catcode `\\12\catcode `\$12\catcode
  `\&12\catcode `\#12\catcode `\^12\catcode `\_12\catcode `\%12\relax}%
\providecommand \@@startlink[1]{}%
\providecommand \@@endlink[0]{}%
\providecommand \url  [0]{\begingroup\@sanitize@url \@url }%
\providecommand \@url [1]{\endgroup\@href {#1}{\urlprefix }}%
\providecommand \urlprefix  [0]{URL }%
\providecommand \Eprint [0]{\href }%
\providecommand \doibase [0]{https://doi.org/}%
\providecommand \selectlanguage [0]{\@gobble}%
\providecommand \bibinfo  [0]{\@secondoftwo}%
\providecommand \bibfield  [0]{\@secondoftwo}%
\providecommand \translation [1]{[#1]}%
\providecommand \BibitemOpen [0]{}%
\providecommand \bibitemStop [0]{}%
\providecommand \bibitemNoStop [0]{.\EOS\space}%
\providecommand \EOS [0]{\spacefactor3000\relax}%
\providecommand \BibitemShut  [1]{\csname bibitem#1\endcsname}%
\let\auto@bib@innerbib\@empty
\bibitem [{\citenamefont {Falk}(2018)}]{Falk2018}%
  \BibitemOpen
  \bibfield  {author} {\bibinfo {author} {\bibfnamefont {K.}~\bibnamefont
  {Falk}},\ }\href {https://doi.org/10.1017/hpl.2018.53} {\bibfield  {journal}
  {\bibinfo  {journal} {High Power Laser Science and Engineering}\ }\textbf
  {\bibinfo {volume} {6}},\ \bibinfo {pages} {e59} (\bibinfo {year} {2018})},\
  \bibinfo {note} {e59}\BibitemShut {NoStop}%
\bibitem [{\citenamefont {Vorberger}\ \emph {et~al.}(2025)\citenamefont
  {Vorberger}, \citenamefont {Graziani}, \citenamefont {Riley}, \citenamefont
  {Baczewski}, \citenamefont {Baraffe}, \citenamefont {Bethkenhagen},
  \citenamefont {Blouin}, \citenamefont {B\"{o}hme}, \citenamefont {Bonitz},
  \citenamefont {Bussmann}, \citenamefont {Casner}, \citenamefont {Cayzac},
  \citenamefont {Celliers}, \citenamefont {Chabrier}, \citenamefont {Chamel},
  \citenamefont {Chapman}, \citenamefont {Chen}, \citenamefont {Cl\'{e}rouin},
  \citenamefont {Collins}, \citenamefont {Coppari}, \citenamefont
  {D\"{o}ppner}, \citenamefont {Dornheim}, \citenamefont {Fletcher},
  \citenamefont {Gericke}, \citenamefont {Glenzer}, \citenamefont {Goncharov},
  \citenamefont {Gregori}, \citenamefont {Hamel}, \citenamefont {Hansen},
  \citenamefont {Hartley}, \citenamefont {Hu}, \citenamefont {Hurricane},
  \citenamefont {Karasiev}, \citenamefont {Kas}, \citenamefont {Kettle},
  \citenamefont {Kluge}, \citenamefont {Knudson}, \citenamefont {Kononov},
  \citenamefont {\'{a}}, \citenamefont {Kraus}, \citenamefont {Kritcher},
  \citenamefont {Malko}, \citenamefont {Massacrier}, \citenamefont {Militzer},
  \citenamefont {Moldabekov}, \citenamefont {Murillo}, \citenamefont {Nagler},
  \citenamefont {Nettelmann}, \citenamefont {Neumayer}, \citenamefont
  {Ofori-Okai}, \citenamefont {Oleynik}, \citenamefont {Preising},
  \citenamefont {Pribram-Jones}, \citenamefont {Ramazanov}, \citenamefont
  {Ravasio}, \citenamefont {Redmer}, \citenamefont {Rethfeld}, \citenamefont
  {Robinson}, \citenamefont {R\"{o}pke}, \citenamefont {Soubiran},
  \citenamefont {Starrett}, \citenamefont {Steinle-Neumann}, \citenamefont
  {Sterne}, \citenamefont {Tanaka}, \citenamefont {Thompson}, \citenamefont
  {Trickey}, \citenamefont {Vinci}, \citenamefont {Vinko}, \citenamefont
  {Wang}, \citenamefont {White}, \citenamefont {White}, \citenamefont
  {Zastrau}, \citenamefont {Zurek},\ and\ \citenamefont
  {Tolias}}]{2025roadmapwarmdensematter}%
  \BibitemOpen
  \bibfield  {author} {\bibinfo {author} {\bibfnamefont {J.}~\bibnamefont
  {Vorberger}}, \bibinfo {author} {\bibfnamefont {F.}~\bibnamefont {Graziani}},
  \bibinfo {author} {\bibfnamefont {D.}~\bibnamefont {Riley}}, \bibinfo
  {author} {\bibfnamefont {A.~D.}\ \bibnamefont {Baczewski}}, \bibinfo {author}
  {\bibfnamefont {I.}~\bibnamefont {Baraffe}}, \bibinfo {author} {\bibfnamefont
  {M.}~\bibnamefont {Bethkenhagen}}, \bibinfo {author} {\bibfnamefont
  {S.}~\bibnamefont {Blouin}}, \bibinfo {author} {\bibfnamefont {M.~P.}\
  \bibnamefont {B\"{o}hme}}, \bibinfo {author} {\bibfnamefont {M.}~\bibnamefont
  {Bonitz}}, \bibinfo {author} {\bibfnamefont {M.}~\bibnamefont {Bussmann}},
  \bibinfo {author} {\bibfnamefont {A.}~\bibnamefont {Casner}}, \bibinfo
  {author} {\bibfnamefont {W.}~\bibnamefont {Cayzac}}, \bibinfo {author}
  {\bibfnamefont {P.}~\bibnamefont {Celliers}}, \bibinfo {author}
  {\bibfnamefont {G.}~\bibnamefont {Chabrier}}, \bibinfo {author}
  {\bibfnamefont {N.}~\bibnamefont {Chamel}}, \bibinfo {author} {\bibfnamefont
  {D.}~\bibnamefont {Chapman}}, \bibinfo {author} {\bibfnamefont
  {M.}~\bibnamefont {Chen}}, \bibinfo {author} {\bibfnamefont {J.}~\bibnamefont
  {Cl\'{e}rouin}}, \bibinfo {author} {\bibfnamefont {G.}~\bibnamefont
  {Collins}}, \bibinfo {author} {\bibfnamefont {F.}~\bibnamefont {Coppari}},
  \bibinfo {author} {\bibfnamefont {T.}~\bibnamefont {D\"{o}ppner}}, \bibinfo
  {author} {\bibfnamefont {T.}~\bibnamefont {Dornheim}}, \bibinfo {author}
  {\bibfnamefont {L.~B.}\ \bibnamefont {Fletcher}}, \bibinfo {author}
  {\bibfnamefont {D.~O.}\ \bibnamefont {Gericke}}, \bibinfo {author}
  {\bibfnamefont {S.}~\bibnamefont {Glenzer}}, \bibinfo {author} {\bibfnamefont
  {A.~F.}\ \bibnamefont {Goncharov}}, \bibinfo {author} {\bibfnamefont
  {G.}~\bibnamefont {Gregori}}, \bibinfo {author} {\bibfnamefont
  {S.}~\bibnamefont {Hamel}}, \bibinfo {author} {\bibfnamefont {S.~B.}\
  \bibnamefont {Hansen}}, \bibinfo {author} {\bibfnamefont {N.~J.}\
  \bibnamefont {Hartley}}, \bibinfo {author} {\bibfnamefont {S.}~\bibnamefont
  {Hu}}, \bibinfo {author} {\bibfnamefont {O.~A.}\ \bibnamefont {Hurricane}},
  \bibinfo {author} {\bibfnamefont {V.~V.}\ \bibnamefont {Karasiev}}, \bibinfo
  {author} {\bibfnamefont {J.~J.}\ \bibnamefont {Kas}}, \bibinfo {author}
  {\bibfnamefont {B.}~\bibnamefont {Kettle}}, \bibinfo {author} {\bibfnamefont
  {T.}~\bibnamefont {Kluge}}, \bibinfo {author} {\bibfnamefont {M.~D.}\
  \bibnamefont {Knudson}}, \bibinfo {author} {\bibfnamefont {A.}~\bibnamefont
  {Kononov}}, \bibinfo {author} {\bibfnamefont {Z.~K.}\ \bibnamefont {\'{a}}},
  \bibinfo {author} {\bibfnamefont {D.}~\bibnamefont {Kraus}}, \bibinfo
  {author} {\bibfnamefont {A.}~\bibnamefont {Kritcher}}, \bibinfo {author}
  {\bibfnamefont {S.}~\bibnamefont {Malko}}, \bibinfo {author} {\bibfnamefont
  {G.}~\bibnamefont {Massacrier}}, \bibinfo {author} {\bibfnamefont
  {B.}~\bibnamefont {Militzer}}, \bibinfo {author} {\bibfnamefont {Z.~A.}\
  \bibnamefont {Moldabekov}}, \bibinfo {author} {\bibfnamefont {M.~S.}\
  \bibnamefont {Murillo}}, \bibinfo {author} {\bibfnamefont {B.}~\bibnamefont
  {Nagler}}, \bibinfo {author} {\bibfnamefont {N.}~\bibnamefont {Nettelmann}},
  \bibinfo {author} {\bibfnamefont {P.}~\bibnamefont {Neumayer}}, \bibinfo
  {author} {\bibfnamefont {B.~K.}\ \bibnamefont {Ofori-Okai}}, \bibinfo
  {author} {\bibfnamefont {I.~I.}\ \bibnamefont {Oleynik}}, \bibinfo {author}
  {\bibfnamefont {M.}~\bibnamefont {Preising}}, \bibinfo {author}
  {\bibfnamefont {A.}~\bibnamefont {Pribram-Jones}}, \bibinfo {author}
  {\bibfnamefont {T.}~\bibnamefont {Ramazanov}}, \bibinfo {author}
  {\bibfnamefont {A.}~\bibnamefont {Ravasio}}, \bibinfo {author} {\bibfnamefont
  {R.}~\bibnamefont {Redmer}}, \bibinfo {author} {\bibfnamefont
  {B.}~\bibnamefont {Rethfeld}}, \bibinfo {author} {\bibfnamefont {A.~P.~L.}\
  \bibnamefont {Robinson}}, \bibinfo {author} {\bibfnamefont {G.}~\bibnamefont
  {R\"{o}pke}}, \bibinfo {author} {\bibfnamefont {F.}~\bibnamefont {Soubiran}},
  \bibinfo {author} {\bibfnamefont {C.~E.}\ \bibnamefont {Starrett}}, \bibinfo
  {author} {\bibfnamefont {G.}~\bibnamefont {Steinle-Neumann}}, \bibinfo
  {author} {\bibfnamefont {P.~A.}\ \bibnamefont {Sterne}}, \bibinfo {author}
  {\bibfnamefont {S.}~\bibnamefont {Tanaka}}, \bibinfo {author} {\bibfnamefont
  {A.~P.}\ \bibnamefont {Thompson}}, \bibinfo {author} {\bibfnamefont {S.~B.}\
  \bibnamefont {Trickey}}, \bibinfo {author} {\bibfnamefont {T.}~\bibnamefont
  {Vinci}}, \bibinfo {author} {\bibfnamefont {S.~M.}\ \bibnamefont {Vinko}},
  \bibinfo {author} {\bibfnamefont {L.}~\bibnamefont {Wang}}, \bibinfo {author}
  {\bibfnamefont {A.~J.}\ \bibnamefont {White}}, \bibinfo {author}
  {\bibfnamefont {T.~G.}\ \bibnamefont {White}}, \bibinfo {author}
  {\bibfnamefont {U.}~\bibnamefont {Zastrau}}, \bibinfo {author} {\bibfnamefont
  {E.}~\bibnamefont {Zurek}},\ and\ \bibinfo {author} {\bibfnamefont
  {P.}~\bibnamefont {Tolias}},\ }\href {https://arxiv.org/abs/2505.02494}
  {\bibinfo {title} {{Roadmap for warm dense matter physics}}} (\bibinfo {year}
  {2025}),\ \Eprint {https://arxiv.org/abs/2505.02494} {arXiv:2505.02494
  [physics.plasm-ph]} \BibitemShut {NoStop}%
\bibitem [{\citenamefont {Militzer}\ \emph {et~al.}(2016)\citenamefont
  {Militzer}, \citenamefont {Soubiran}, \citenamefont {Wahl},\ and\
  \citenamefont {Hubbard}}]{Militzer16}%
  \BibitemOpen
  \bibfield  {author} {\bibinfo {author} {\bibfnamefont {B.}~\bibnamefont
  {Militzer}}, \bibinfo {author} {\bibfnamefont {F.}~\bibnamefont {Soubiran}},
  \bibinfo {author} {\bibfnamefont {S.~M.}\ \bibnamefont {Wahl}},\ and\
  \bibinfo {author} {\bibfnamefont {W.}~\bibnamefont {Hubbard}},\ }\href
  {https://doi.org/https://doi.org/10.1002/2016JE005080} {\bibfield  {journal}
  {\bibinfo  {journal} {Journal of Geophysical Research: Planets}\ }\textbf
  {\bibinfo {volume} {121}},\ \bibinfo {pages} {1552} (\bibinfo {year}
  {2016})}\BibitemShut {NoStop}%
\bibitem [{\citenamefont {Helled}\ \emph {et~al.}(2020)\citenamefont {Helled},
  \citenamefont {Mazzola},\ and\ \citenamefont {Redmer}}]{Helled2020}%
  \BibitemOpen
  \bibfield  {author} {\bibinfo {author} {\bibfnamefont {R.}~\bibnamefont
  {Helled}}, \bibinfo {author} {\bibfnamefont {G.}~\bibnamefont {Mazzola}},\
  and\ \bibinfo {author} {\bibfnamefont {R.}~\bibnamefont {Redmer}},\ }\href
  {https://doi.org/10.1038/s42254-020-0223-3} {\bibfield  {journal} {\bibinfo
  {journal} {Nature Reviews Physics}\ }\textbf {\bibinfo {volume} {2}},\
  \bibinfo {pages} {562} (\bibinfo {year} {2020})}\BibitemShut {NoStop}%
\bibitem [{\citenamefont {Saumon}\ \emph {et~al.}(2022)\citenamefont {Saumon},
  \citenamefont {Blouin},\ and\ \citenamefont {Tremblay}}]{SAUMON20221}%
  \BibitemOpen
  \bibfield  {author} {\bibinfo {author} {\bibfnamefont {D.}~\bibnamefont
  {Saumon}}, \bibinfo {author} {\bibfnamefont {S.}~\bibnamefont {Blouin}},\
  and\ \bibinfo {author} {\bibfnamefont {P.-E.}\ \bibnamefont {Tremblay}},\
  }\href {https://doi.org/https://doi.org/10.1016/j.physrep.2022.09.001}
  {\bibfield  {journal} {\bibinfo  {journal} {Physics Reports}\ }\textbf
  {\bibinfo {volume} {988}},\ \bibinfo {pages} {1} (\bibinfo {year} {2022})},\
  \bibinfo {note} {current Challenges in the Physics of White Dwarf
  Stars}\BibitemShut {NoStop}%
\bibitem [{\citenamefont {Breuer}\ and\ \citenamefont {Spohn}(2023)}]{Doris23}%
  \BibitemOpen
  \bibfield  {author} {\bibinfo {author} {\bibfnamefont {D.}~\bibnamefont
  {Breuer}}\ and\ \bibinfo {author} {\bibfnamefont {T.}~\bibnamefont {Spohn}},\
  }\href {https://doi.org/10.1093/acrefore/9780190647926.013.28} {\bibinfo
  {title} {Terrestrial planets: Interior structure, dynamics, and evolution}}
  (\bibinfo {year} {2023})\BibitemShut {NoStop}%
\bibitem [{\citenamefont {Roy}\ \emph {et~al.}(2024)\citenamefont {Roy},
  \citenamefont {Bergermann}, \citenamefont {Bethkenhagen},\ and\ \citenamefont
  {Redmer}}]{Roy24}%
  \BibitemOpen
  \bibfield  {author} {\bibinfo {author} {\bibfnamefont {A.~J.}\ \bibnamefont
  {Roy}}, \bibinfo {author} {\bibfnamefont {A.}~\bibnamefont {Bergermann}},
  \bibinfo {author} {\bibfnamefont {M.}~\bibnamefont {Bethkenhagen}},\ and\
  \bibinfo {author} {\bibfnamefont {R.}~\bibnamefont {Redmer}},\ }\href
  {https://doi.org/10.1039/D4CP00058G} {\bibfield  {journal} {\bibinfo
  {journal} {Phys. Chem. Chem. Phys.}\ }\textbf {\bibinfo {volume} {26}},\
  \bibinfo {pages} {14374} (\bibinfo {year} {2024})}\BibitemShut {NoStop}%
\bibitem [{\citenamefont {Roy}\ \emph {et~al.}(2025)\citenamefont {Roy},
  \citenamefont {Bethkenhagen}, \citenamefont {Bergermann},\ and\ \citenamefont
  {Redmer}}]{Roy2025}%
  \BibitemOpen
  \bibfield  {author} {\bibinfo {author} {\bibfnamefont {A.~J.}\ \bibnamefont
  {Roy}}, \bibinfo {author} {\bibfnamefont {M.}~\bibnamefont {Bethkenhagen}},
  \bibinfo {author} {\bibfnamefont {A.}~\bibnamefont {Bergermann}},\ and\
  \bibinfo {author} {\bibfnamefont {R.}~\bibnamefont {Redmer}},\ }\href
  {https://doi.org/10.3847/1538-4357/adf5be} {\bibfield  {journal} {\bibinfo
  {journal} {The Astrophysical Journal}\ }\textbf {\bibinfo {volume} {990}},\
  \bibinfo {pages} {212} (\bibinfo {year} {2025})}\BibitemShut {NoStop}%
\bibitem [{Eur(2025)}]{Europa}%
  \BibitemOpen
  \bibfield  {journal} {\bibinfo  {journal} {Nature Communications}\ }\textbf
  {\bibinfo {volume} {16}},\ \href {https://doi.org/10.1038/s41467-025-59094-6}
  {10.1038/s41467-025-59094-6} (\bibinfo {year} {2025})\BibitemShut {NoStop}%
\bibitem [{\citenamefont {Patel}\ \emph {et~al.}(2020)\citenamefont {Patel},
  \citenamefont {Springer}, \citenamefont {Weber}, \citenamefont {Jarrott},
  \citenamefont {Hurricane}, \citenamefont {Bachmann}, \citenamefont {Baker},
  \citenamefont {Berzak~Hopkins}, \citenamefont {Callahan}, \citenamefont
  {Casey}, \citenamefont {Cerjan}, \citenamefont {Clark}, \citenamefont
  {Dewald}, \citenamefont {Divol}, \citenamefont {D\"{o}ppner}, \citenamefont
  {Field}, \citenamefont {Fittinghoff}, \citenamefont {Gaffney}, \citenamefont
  {Geppert-Kleinrath}, \citenamefont {Grim}, \citenamefont {Hartouni},
  \citenamefont {Hatarik}, \citenamefont {Hinkel}, \citenamefont {Hohenberger},
  \citenamefont {Humbird}, \citenamefont {Izumi}, \citenamefont {Jones},
  \citenamefont {Khan}, \citenamefont {Kritcher}, \citenamefont {Kruse},
  \citenamefont {Landen}, \citenamefont {Le~Pape}, \citenamefont {Ma},
  \citenamefont {MacLaren}, \citenamefont {MacPhee}, \citenamefont {Masse},
  \citenamefont {Meezan}, \citenamefont {Milovich}, \citenamefont {Nora},
  \citenamefont {Pak}, \citenamefont {Peterson}, \citenamefont {Ralph},
  \citenamefont {Robey}, \citenamefont {Salmonson}, \citenamefont {Smalyuk},
  \citenamefont {Spears}, \citenamefont {Thomas}, \citenamefont {Volegov},
  \citenamefont {Zylstra},\ and\ \citenamefont {Edwards}}]{ICF2020}%
  \BibitemOpen
  \bibfield  {author} {\bibinfo {author} {\bibfnamefont {P.~K.}\ \bibnamefont
  {Patel}}, \bibinfo {author} {\bibfnamefont {P.~T.}\ \bibnamefont {Springer}},
  \bibinfo {author} {\bibfnamefont {C.~R.}\ \bibnamefont {Weber}}, \bibinfo
  {author} {\bibfnamefont {L.~C.}\ \bibnamefont {Jarrott}}, \bibinfo {author}
  {\bibfnamefont {O.~A.}\ \bibnamefont {Hurricane}}, \bibinfo {author}
  {\bibfnamefont {B.}~\bibnamefont {Bachmann}}, \bibinfo {author}
  {\bibfnamefont {K.~L.}\ \bibnamefont {Baker}}, \bibinfo {author}
  {\bibfnamefont {L.~F.}\ \bibnamefont {Berzak~Hopkins}}, \bibinfo {author}
  {\bibfnamefont {D.~A.}\ \bibnamefont {Callahan}}, \bibinfo {author}
  {\bibfnamefont {D.~T.}\ \bibnamefont {Casey}}, \bibinfo {author}
  {\bibfnamefont {C.~J.}\ \bibnamefont {Cerjan}}, \bibinfo {author}
  {\bibfnamefont {D.~S.}\ \bibnamefont {Clark}}, \bibinfo {author}
  {\bibfnamefont {E.~L.}\ \bibnamefont {Dewald}}, \bibinfo {author}
  {\bibfnamefont {L.}~\bibnamefont {Divol}}, \bibinfo {author} {\bibfnamefont
  {T.}~\bibnamefont {D\"{o}ppner}}, \bibinfo {author} {\bibfnamefont {J.~E.}\
  \bibnamefont {Field}}, \bibinfo {author} {\bibfnamefont {D.}~\bibnamefont
  {Fittinghoff}}, \bibinfo {author} {\bibfnamefont {J.}~\bibnamefont
  {Gaffney}}, \bibinfo {author} {\bibfnamefont {V.}~\bibnamefont
  {Geppert-Kleinrath}}, \bibinfo {author} {\bibfnamefont {G.~P.}\ \bibnamefont
  {Grim}}, \bibinfo {author} {\bibfnamefont {E.~P.}\ \bibnamefont {Hartouni}},
  \bibinfo {author} {\bibfnamefont {R.}~\bibnamefont {Hatarik}}, \bibinfo
  {author} {\bibfnamefont {D.~E.}\ \bibnamefont {Hinkel}}, \bibinfo {author}
  {\bibfnamefont {M.}~\bibnamefont {Hohenberger}}, \bibinfo {author}
  {\bibfnamefont {K.}~\bibnamefont {Humbird}}, \bibinfo {author} {\bibfnamefont
  {N.}~\bibnamefont {Izumi}}, \bibinfo {author} {\bibfnamefont {O.~S.}\
  \bibnamefont {Jones}}, \bibinfo {author} {\bibfnamefont {S.~F.}\ \bibnamefont
  {Khan}}, \bibinfo {author} {\bibfnamefont {A.~L.}\ \bibnamefont {Kritcher}},
  \bibinfo {author} {\bibfnamefont {M.}~\bibnamefont {Kruse}}, \bibinfo
  {author} {\bibfnamefont {O.~L.}\ \bibnamefont {Landen}}, \bibinfo {author}
  {\bibfnamefont {S.}~\bibnamefont {Le~Pape}}, \bibinfo {author} {\bibfnamefont
  {T.}~\bibnamefont {Ma}}, \bibinfo {author} {\bibfnamefont {S.~A.}\
  \bibnamefont {MacLaren}}, \bibinfo {author} {\bibfnamefont {A.~G.}\
  \bibnamefont {MacPhee}}, \bibinfo {author} {\bibfnamefont {L.~P.}\
  \bibnamefont {Masse}}, \bibinfo {author} {\bibfnamefont {N.~B.}\ \bibnamefont
  {Meezan}}, \bibinfo {author} {\bibfnamefont {J.~L.}\ \bibnamefont
  {Milovich}}, \bibinfo {author} {\bibfnamefont {R.}~\bibnamefont {Nora}},
  \bibinfo {author} {\bibfnamefont {A.}~\bibnamefont {Pak}}, \bibinfo {author}
  {\bibfnamefont {J.~L.}\ \bibnamefont {Peterson}}, \bibinfo {author}
  {\bibfnamefont {J.}~\bibnamefont {Ralph}}, \bibinfo {author} {\bibfnamefont
  {H.~F.}\ \bibnamefont {Robey}}, \bibinfo {author} {\bibfnamefont {J.~D.}\
  \bibnamefont {Salmonson}}, \bibinfo {author} {\bibfnamefont {V.~A.}\
  \bibnamefont {Smalyuk}}, \bibinfo {author} {\bibfnamefont {B.~K.}\
  \bibnamefont {Spears}}, \bibinfo {author} {\bibfnamefont {C.~A.}\
  \bibnamefont {Thomas}}, \bibinfo {author} {\bibfnamefont {P.~L.}\
  \bibnamefont {Volegov}}, \bibinfo {author} {\bibfnamefont {A.}~\bibnamefont
  {Zylstra}},\ and\ \bibinfo {author} {\bibfnamefont {M.~J.}\ \bibnamefont
  {Edwards}},\ }\href {https://doi.org/10.1063/5.0003298} {\bibfield  {journal}
  {\bibinfo  {journal} {Physics of Plasmas}\ }\textbf {\bibinfo {volume}
  {27}},\ \bibinfo {pages} {050901} (\bibinfo {year} {2020})}\BibitemShut
  {NoStop}%
\bibitem [{\citenamefont {Olson}\ \emph {et~al.}(2021)\citenamefont {Olson},
  \citenamefont {Schmitt}, \citenamefont {Haines}, \citenamefont {Kemp},
  \citenamefont {Yeamans}, \citenamefont {Blue}, \citenamefont {Schmidt},
  \citenamefont {Haid}, \citenamefont {Farrell}, \citenamefont {Bradley},
  \citenamefont {Robey},\ and\ \citenamefont {Leeper}}]{ICF2021}%
  \BibitemOpen
  \bibfield  {author} {\bibinfo {author} {\bibfnamefont {R.~E.}\ \bibnamefont
  {Olson}}, \bibinfo {author} {\bibfnamefont {M.~J.}\ \bibnamefont {Schmitt}},
  \bibinfo {author} {\bibfnamefont {B.~M.}\ \bibnamefont {Haines}}, \bibinfo
  {author} {\bibfnamefont {G.~E.}\ \bibnamefont {Kemp}}, \bibinfo {author}
  {\bibfnamefont {C.~B.}\ \bibnamefont {Yeamans}}, \bibinfo {author}
  {\bibfnamefont {B.~E.}\ \bibnamefont {Blue}}, \bibinfo {author}
  {\bibfnamefont {D.~W.}\ \bibnamefont {Schmidt}}, \bibinfo {author}
  {\bibfnamefont {A.}~\bibnamefont {Haid}}, \bibinfo {author} {\bibfnamefont
  {M.}~\bibnamefont {Farrell}}, \bibinfo {author} {\bibfnamefont {P.~A.}\
  \bibnamefont {Bradley}}, \bibinfo {author} {\bibfnamefont {H.~F.}\
  \bibnamefont {Robey}},\ and\ \bibinfo {author} {\bibfnamefont {R.~J.}\
  \bibnamefont {Leeper}},\ }\href {https://doi.org/10.1063/5.0062590}
  {\bibfield  {journal} {\bibinfo  {journal} {Physics of Plasmas}\ }\textbf
  {\bibinfo {volume} {28}},\ \bibinfo {pages} {122704} (\bibinfo {year}
  {2021})}\BibitemShut {NoStop}%
\bibitem [{\citenamefont {Grabowski}\ \emph {et~al.}(2020)\citenamefont
  {Grabowski}, \citenamefont {Hansen}, \citenamefont {Murillo}, \citenamefont
  {Stanton}, \citenamefont {Graziani}, \citenamefont {Zylstra}, \citenamefont
  {Baalrud}, \citenamefont {Arnault}, \citenamefont {Baczewski}, \citenamefont
  {Benedict}, \citenamefont {Blancard}, \citenamefont {Čertík}, \citenamefont
  {Clérouin}, \citenamefont {Collins}, \citenamefont {Copeland}, \citenamefont
  {Correa}, \citenamefont {Dai}, \citenamefont {Daligault}, \citenamefont
  {Desjarlais}, \citenamefont {Dharma-wardana}, \citenamefont {Faussurier},
  \citenamefont {Haack}, \citenamefont {Haxhimali}, \citenamefont
  {Hayes-Sterbenz}, \citenamefont {Hou}, \citenamefont {Hu}, \citenamefont
  {Jensen}, \citenamefont {Jungman}, \citenamefont {Kagan}, \citenamefont
  {Kang}, \citenamefont {Kress}, \citenamefont {Ma}, \citenamefont {Marciante},
  \citenamefont {Meyer}, \citenamefont {Rudd}, \citenamefont {Saumon},
  \citenamefont {Shulenburger}, \citenamefont {Singleton}, \citenamefont
  {Sjostrom}, \citenamefont {Stanek}, \citenamefont {Starrett}, \citenamefont
  {Ticknor}, \citenamefont {Valaitis}, \citenamefont {Venzke},\ and\
  \citenamefont {White}}]{Grabowski20}%
  \BibitemOpen
  \bibfield  {author} {\bibinfo {author} {\bibfnamefont {P.}~\bibnamefont
  {Grabowski}}, \bibinfo {author} {\bibfnamefont {S.}~\bibnamefont {Hansen}},
  \bibinfo {author} {\bibfnamefont {M.}~\bibnamefont {Murillo}}, \bibinfo
  {author} {\bibfnamefont {L.}~\bibnamefont {Stanton}}, \bibinfo {author}
  {\bibfnamefont {F.}~\bibnamefont {Graziani}}, \bibinfo {author}
  {\bibfnamefont {A.}~\bibnamefont {Zylstra}}, \bibinfo {author} {\bibfnamefont
  {S.}~\bibnamefont {Baalrud}}, \bibinfo {author} {\bibfnamefont
  {P.}~\bibnamefont {Arnault}}, \bibinfo {author} {\bibfnamefont
  {A.}~\bibnamefont {Baczewski}}, \bibinfo {author} {\bibfnamefont
  {L.}~\bibnamefont {Benedict}}, \bibinfo {author} {\bibfnamefont
  {C.}~\bibnamefont {Blancard}}, \bibinfo {author} {\bibfnamefont
  {O.}~\bibnamefont {Čertík}}, \bibinfo {author} {\bibfnamefont
  {J.}~\bibnamefont {Clérouin}}, \bibinfo {author} {\bibfnamefont
  {L.}~\bibnamefont {Collins}}, \bibinfo {author} {\bibfnamefont
  {S.}~\bibnamefont {Copeland}}, \bibinfo {author} {\bibfnamefont
  {A.}~\bibnamefont {Correa}}, \bibinfo {author} {\bibfnamefont
  {J.}~\bibnamefont {Dai}}, \bibinfo {author} {\bibfnamefont {J.}~\bibnamefont
  {Daligault}}, \bibinfo {author} {\bibfnamefont {M.}~\bibnamefont
  {Desjarlais}}, \bibinfo {author} {\bibfnamefont {M.}~\bibnamefont
  {Dharma-wardana}}, \bibinfo {author} {\bibfnamefont {G.}~\bibnamefont
  {Faussurier}}, \bibinfo {author} {\bibfnamefont {J.}~\bibnamefont {Haack}},
  \bibinfo {author} {\bibfnamefont {T.}~\bibnamefont {Haxhimali}}, \bibinfo
  {author} {\bibfnamefont {A.}~\bibnamefont {Hayes-Sterbenz}}, \bibinfo
  {author} {\bibfnamefont {Y.}~\bibnamefont {Hou}}, \bibinfo {author}
  {\bibfnamefont {S.}~\bibnamefont {Hu}}, \bibinfo {author} {\bibfnamefont
  {D.}~\bibnamefont {Jensen}}, \bibinfo {author} {\bibfnamefont
  {G.}~\bibnamefont {Jungman}}, \bibinfo {author} {\bibfnamefont
  {G.}~\bibnamefont {Kagan}}, \bibinfo {author} {\bibfnamefont
  {D.}~\bibnamefont {Kang}}, \bibinfo {author} {\bibfnamefont {J.}~\bibnamefont
  {Kress}}, \bibinfo {author} {\bibfnamefont {Q.}~\bibnamefont {Ma}}, \bibinfo
  {author} {\bibfnamefont {M.}~\bibnamefont {Marciante}}, \bibinfo {author}
  {\bibfnamefont {E.}~\bibnamefont {Meyer}}, \bibinfo {author} {\bibfnamefont
  {R.}~\bibnamefont {Rudd}}, \bibinfo {author} {\bibfnamefont {D.}~\bibnamefont
  {Saumon}}, \bibinfo {author} {\bibfnamefont {L.}~\bibnamefont
  {Shulenburger}}, \bibinfo {author} {\bibfnamefont {R.}~\bibnamefont
  {Singleton}}, \bibinfo {author} {\bibfnamefont {T.}~\bibnamefont {Sjostrom}},
  \bibinfo {author} {\bibfnamefont {L.}~\bibnamefont {Stanek}}, \bibinfo
  {author} {\bibfnamefont {C.}~\bibnamefont {Starrett}}, \bibinfo {author}
  {\bibfnamefont {C.}~\bibnamefont {Ticknor}}, \bibinfo {author} {\bibfnamefont
  {S.}~\bibnamefont {Valaitis}}, \bibinfo {author} {\bibfnamefont
  {J.}~\bibnamefont {Venzke}},\ and\ \bibinfo {author} {\bibfnamefont
  {A.}~\bibnamefont {White}},\ }\href
  {https://doi.org/https://doi.org/10.1016/j.hedp.2020.100905} {\bibfield
  {journal} {\bibinfo  {journal} {High Energy Density Physics}\ }\textbf
  {\bibinfo {volume} {37}},\ \bibinfo {pages} {100905} (\bibinfo {year}
  {2020})}\BibitemShut {NoStop}%
\bibitem [{\citenamefont {Stanek}\ \emph {et~al.}(2024)\citenamefont {Stanek},
  \citenamefont {Kononov}, \citenamefont {Hansen}, \citenamefont {Haines},
  \citenamefont {Hu}, \citenamefont {Knapp}, \citenamefont {Murillo},
  \citenamefont {Stanton}, \citenamefont {Whitley}, \citenamefont {Baalrud},
  \citenamefont {Babati}, \citenamefont {Baczewski}, \citenamefont
  {Bethkenhagen}, \citenamefont {Blanchet}, \citenamefont {Clay}, \citenamefont
  {Cochrane}, \citenamefont {Collins}, \citenamefont {Dumi}, \citenamefont
  {Faussurier}, \citenamefont {French}, \citenamefont {Johnson}, \citenamefont
  {Karasiev}, \citenamefont {Kumar}, \citenamefont {Lentz}, \citenamefont
  {Melton}, \citenamefont {Nichols}, \citenamefont {Petrov}, \citenamefont
  {Recoules}, \citenamefont {Redmer}, \citenamefont {R\"{o}pke}, \citenamefont
  {Sch\"{o}rner}, \citenamefont {Shaffer}, \citenamefont {Sharma},
  \citenamefont {Silvestri}, \citenamefont {Soubiran}, \citenamefont
  {Suryanarayana}, \citenamefont {Tacu}, \citenamefont {Townsend},\ and\
  \citenamefont {White}}]{Stanek24}%
  \BibitemOpen
  \bibfield  {author} {\bibinfo {author} {\bibfnamefont {L.~J.}\ \bibnamefont
  {Stanek}}, \bibinfo {author} {\bibfnamefont {A.}~\bibnamefont {Kononov}},
  \bibinfo {author} {\bibfnamefont {S.~B.}\ \bibnamefont {Hansen}}, \bibinfo
  {author} {\bibfnamefont {B.~M.}\ \bibnamefont {Haines}}, \bibinfo {author}
  {\bibfnamefont {S.~X.}\ \bibnamefont {Hu}}, \bibinfo {author} {\bibfnamefont
  {P.~F.}\ \bibnamefont {Knapp}}, \bibinfo {author} {\bibfnamefont {M.~S.}\
  \bibnamefont {Murillo}}, \bibinfo {author} {\bibfnamefont {L.~G.}\
  \bibnamefont {Stanton}}, \bibinfo {author} {\bibfnamefont {H.~D.}\
  \bibnamefont {Whitley}}, \bibinfo {author} {\bibfnamefont {S.~D.}\
  \bibnamefont {Baalrud}}, \bibinfo {author} {\bibfnamefont {L.~J.}\
  \bibnamefont {Babati}}, \bibinfo {author} {\bibfnamefont {A.~D.}\
  \bibnamefont {Baczewski}}, \bibinfo {author} {\bibfnamefont {M.}~\bibnamefont
  {Bethkenhagen}}, \bibinfo {author} {\bibfnamefont {A.}~\bibnamefont
  {Blanchet}}, \bibinfo {author} {\bibfnamefont {I.}~\bibnamefont {Clay},
  \bibfnamefont {Raymond~C.}}, \bibinfo {author} {\bibfnamefont {K.~R.}\
  \bibnamefont {Cochrane}}, \bibinfo {author} {\bibfnamefont {L.~A.}\
  \bibnamefont {Collins}}, \bibinfo {author} {\bibfnamefont {A.}~\bibnamefont
  {Dumi}}, \bibinfo {author} {\bibfnamefont {G.}~\bibnamefont {Faussurier}},
  \bibinfo {author} {\bibfnamefont {M.}~\bibnamefont {French}}, \bibinfo
  {author} {\bibfnamefont {Z.~A.}\ \bibnamefont {Johnson}}, \bibinfo {author}
  {\bibfnamefont {V.~V.}\ \bibnamefont {Karasiev}}, \bibinfo {author}
  {\bibfnamefont {S.}~\bibnamefont {Kumar}}, \bibinfo {author} {\bibfnamefont
  {M.~K.}\ \bibnamefont {Lentz}}, \bibinfo {author} {\bibfnamefont {C.~A.}\
  \bibnamefont {Melton}}, \bibinfo {author} {\bibfnamefont {K.~A.}\
  \bibnamefont {Nichols}}, \bibinfo {author} {\bibfnamefont {G.~M.}\
  \bibnamefont {Petrov}}, \bibinfo {author} {\bibfnamefont {V.}~\bibnamefont
  {Recoules}}, \bibinfo {author} {\bibfnamefont {R.}~\bibnamefont {Redmer}},
  \bibinfo {author} {\bibfnamefont {G.}~\bibnamefont {R\"{o}pke}}, \bibinfo
  {author} {\bibfnamefont {M.}~\bibnamefont {Sch\"{o}rner}}, \bibinfo {author}
  {\bibfnamefont {N.~R.}\ \bibnamefont {Shaffer}}, \bibinfo {author}
  {\bibfnamefont {V.}~\bibnamefont {Sharma}}, \bibinfo {author} {\bibfnamefont
  {L.~G.}\ \bibnamefont {Silvestri}}, \bibinfo {author} {\bibfnamefont
  {F.}~\bibnamefont {Soubiran}}, \bibinfo {author} {\bibfnamefont
  {P.}~\bibnamefont {Suryanarayana}}, \bibinfo {author} {\bibfnamefont
  {M.}~\bibnamefont {Tacu}}, \bibinfo {author} {\bibfnamefont {J.~P.}\
  \bibnamefont {Townsend}},\ and\ \bibinfo {author} {\bibfnamefont {A.~J.}\
  \bibnamefont {White}},\ }\href {https://doi.org/10.1063/5.0198155} {\bibfield
   {journal} {\bibinfo  {journal} {Physics of Plasmas}\ }\textbf {\bibinfo
  {volume} {31}},\ \bibinfo {pages} {052104} (\bibinfo {year}
  {2024})}\BibitemShut {NoStop}%
\bibitem [{\citenamefont {Hurricane}\ \emph {et~al.}(2023)\citenamefont
  {Hurricane}, \citenamefont {Patel}, \citenamefont {Betti}, \citenamefont
  {Froula}, \citenamefont {Regan}, \citenamefont {Slutz}, \citenamefont
  {Gomez},\ and\ \citenamefont {Sweeney}}]{Hurricane23}%
  \BibitemOpen
  \bibfield  {author} {\bibinfo {author} {\bibfnamefont {O.~A.}\ \bibnamefont
  {Hurricane}}, \bibinfo {author} {\bibfnamefont {P.~K.}\ \bibnamefont
  {Patel}}, \bibinfo {author} {\bibfnamefont {R.}~\bibnamefont {Betti}},
  \bibinfo {author} {\bibfnamefont {D.~H.}\ \bibnamefont {Froula}}, \bibinfo
  {author} {\bibfnamefont {S.~P.}\ \bibnamefont {Regan}}, \bibinfo {author}
  {\bibfnamefont {S.~A.}\ \bibnamefont {Slutz}}, \bibinfo {author}
  {\bibfnamefont {M.~R.}\ \bibnamefont {Gomez}},\ and\ \bibinfo {author}
  {\bibfnamefont {M.~A.}\ \bibnamefont {Sweeney}},\ }\href
  {https://doi.org/10.1103/RevModPhys.95.025005} {\bibfield  {journal}
  {\bibinfo  {journal} {Rev. Mod. Phys.}\ }\textbf {\bibinfo {volume} {95}},\
  \bibinfo {pages} {025005} (\bibinfo {year} {2023})}\BibitemShut {NoStop}%
\bibitem [{\citenamefont {Hu}\ \emph {et~al.}(2024)\citenamefont {Hu},
  \citenamefont {Nichols}, \citenamefont {Shaffer}, \citenamefont {Arnold},
  \citenamefont {White}, \citenamefont {Collins}, \citenamefont {Karasiev},
  \citenamefont {Zhang}, \citenamefont {Goncharov}, \citenamefont {Shah},
  \citenamefont {Mihaylov}, \citenamefont {Jiang},\ and\ \citenamefont
  {Ping}}]{Hu24}%
  \BibitemOpen
  \bibfield  {author} {\bibinfo {author} {\bibfnamefont {S.~X.}\ \bibnamefont
  {Hu}}, \bibinfo {author} {\bibfnamefont {K.~A.}\ \bibnamefont {Nichols}},
  \bibinfo {author} {\bibfnamefont {N.~R.}\ \bibnamefont {Shaffer}}, \bibinfo
  {author} {\bibfnamefont {B.}~\bibnamefont {Arnold}}, \bibinfo {author}
  {\bibfnamefont {A.~J.}\ \bibnamefont {White}}, \bibinfo {author}
  {\bibfnamefont {L.~A.}\ \bibnamefont {Collins}}, \bibinfo {author}
  {\bibfnamefont {V.~V.}\ \bibnamefont {Karasiev}}, \bibinfo {author}
  {\bibfnamefont {S.}~\bibnamefont {Zhang}}, \bibinfo {author} {\bibfnamefont
  {V.~N.}\ \bibnamefont {Goncharov}}, \bibinfo {author} {\bibfnamefont {R.~C.}\
  \bibnamefont {Shah}}, \bibinfo {author} {\bibfnamefont {D.~I.}\ \bibnamefont
  {Mihaylov}}, \bibinfo {author} {\bibfnamefont {S.}~\bibnamefont {Jiang}},\
  and\ \bibinfo {author} {\bibfnamefont {Y.}~\bibnamefont {Ping}},\ }\href
  {https://doi.org/10.1063/5.0197969} {\bibfield  {journal} {\bibinfo
  {journal} {Physics of Plasmas}\ }\textbf {\bibinfo {volume} {31}},\ \bibinfo
  {pages} {040501} (\bibinfo {year} {2024})}\BibitemShut {NoStop}%
\bibitem [{\citenamefont {Haines}(2024)}]{Haines24}%
  \BibitemOpen
  \bibfield  {author} {\bibinfo {author} {\bibfnamefont {B.~M.}\ \bibnamefont
  {Haines}},\ }\href {https://doi.org/10.1063/5.0197128} {\bibfield  {journal}
  {\bibinfo  {journal} {Physics of Plasmas}\ }\textbf {\bibinfo {volume}
  {31}},\ \bibinfo {pages} {050501} (\bibinfo {year} {2024})}\BibitemShut
  {NoStop}%
\bibitem [{\citenamefont {Allen}\ \emph {et~al.}(2025)\citenamefont {Allen},
  \citenamefont {Oliver}, \citenamefont {Gericke}, \citenamefont {Brouwer},
  \citenamefont {Divol}, \citenamefont {Kemp}, \citenamefont {Landen},
  \citenamefont {Morrison}, \citenamefont {Ping}, \citenamefont
  {Sch{\"o}lmerich}, \citenamefont {Shaffer}, \citenamefont {Spindloe},
  \citenamefont {Sterne}, \citenamefont {Theobald}, \citenamefont
  {D{\"o}ppner},\ and\ \citenamefont {White}}]{Allen2025}%
  \BibitemOpen
  \bibfield  {author} {\bibinfo {author} {\bibfnamefont {C.~H.}\ \bibnamefont
  {Allen}}, \bibinfo {author} {\bibfnamefont {M.}~\bibnamefont {Oliver}},
  \bibinfo {author} {\bibfnamefont {D.~O.}\ \bibnamefont {Gericke}}, \bibinfo
  {author} {\bibfnamefont {N.}~\bibnamefont {Brouwer}}, \bibinfo {author}
  {\bibfnamefont {L.}~\bibnamefont {Divol}}, \bibinfo {author} {\bibfnamefont
  {G.~E.}\ \bibnamefont {Kemp}}, \bibinfo {author} {\bibfnamefont {O.~L.}\
  \bibnamefont {Landen}}, \bibinfo {author} {\bibfnamefont {L.}~\bibnamefont
  {Morrison}}, \bibinfo {author} {\bibfnamefont {Y.}~\bibnamefont {Ping}},
  \bibinfo {author} {\bibfnamefont {M.~O.}\ \bibnamefont {Sch{\"o}lmerich}},
  \bibinfo {author} {\bibfnamefont {N.}~\bibnamefont {Shaffer}}, \bibinfo
  {author} {\bibfnamefont {C.}~\bibnamefont {Spindloe}}, \bibinfo {author}
  {\bibfnamefont {P.~A.}\ \bibnamefont {Sterne}}, \bibinfo {author}
  {\bibfnamefont {W.~R.}\ \bibnamefont {Theobald}}, \bibinfo {author}
  {\bibfnamefont {T.}~\bibnamefont {D{\"o}ppner}},\ and\ \bibinfo {author}
  {\bibfnamefont {T.~G.}\ \bibnamefont {White}},\ }\href
  {https://doi.org/10.1038/s41467-025-56051-1} {\bibfield  {journal} {\bibinfo
  {journal} {Nature Communications}\ }\textbf {\bibinfo {volume} {16}},\
  \bibinfo {pages} {1983} (\bibinfo {year} {2025})}\BibitemShut {NoStop}%
\bibitem [{\citenamefont {White}\ \emph {et~al.}(2025)\citenamefont {White},
  \citenamefont {Griffin}, \citenamefont {Haden}, \citenamefont {Lee},
  \citenamefont {Galtier}, \citenamefont {Cunningham}, \citenamefont
  {Khaghani}, \citenamefont {Descamps}, \citenamefont {Wollenweber},
  \citenamefont {Armentrout}, \citenamefont {Convery}, \citenamefont {Appel},
  \citenamefont {Fletcher}, \citenamefont {Goede}, \citenamefont {Hastings},
  \citenamefont {Iratcabal}, \citenamefont {McBride}, \citenamefont {Molina},
  \citenamefont {Monaco}, \citenamefont {Morrison}, \citenamefont {Stramel},
  \citenamefont {Yunus}, \citenamefont {Zastrau}, \citenamefont {Glenzer},
  \citenamefont {Gregori}, \citenamefont {Gericke},\ and\ \citenamefont
  {Nagler}}]{TWhite2025}%
  \BibitemOpen
  \bibfield  {author} {\bibinfo {author} {\bibfnamefont {T.~G.}\ \bibnamefont
  {White}}, \bibinfo {author} {\bibfnamefont {T.~D.}\ \bibnamefont {Griffin}},
  \bibinfo {author} {\bibfnamefont {D.}~\bibnamefont {Haden}}, \bibinfo
  {author} {\bibfnamefont {H.~J.}\ \bibnamefont {Lee}}, \bibinfo {author}
  {\bibfnamefont {E.}~\bibnamefont {Galtier}}, \bibinfo {author} {\bibfnamefont
  {E.}~\bibnamefont {Cunningham}}, \bibinfo {author} {\bibfnamefont
  {D.}~\bibnamefont {Khaghani}}, \bibinfo {author} {\bibfnamefont
  {A.}~\bibnamefont {Descamps}}, \bibinfo {author} {\bibfnamefont
  {L.}~\bibnamefont {Wollenweber}}, \bibinfo {author} {\bibfnamefont
  {B.}~\bibnamefont {Armentrout}}, \bibinfo {author} {\bibfnamefont
  {C.}~\bibnamefont {Convery}}, \bibinfo {author} {\bibfnamefont
  {K.}~\bibnamefont {Appel}}, \bibinfo {author} {\bibfnamefont {L.~B.}\
  \bibnamefont {Fletcher}}, \bibinfo {author} {\bibfnamefont {S.}~\bibnamefont
  {Goede}}, \bibinfo {author} {\bibfnamefont {J.~B.}\ \bibnamefont {Hastings}},
  \bibinfo {author} {\bibfnamefont {J.}~\bibnamefont {Iratcabal}}, \bibinfo
  {author} {\bibfnamefont {E.~E.}\ \bibnamefont {McBride}}, \bibinfo {author}
  {\bibfnamefont {J.}~\bibnamefont {Molina}}, \bibinfo {author} {\bibfnamefont
  {G.}~\bibnamefont {Monaco}}, \bibinfo {author} {\bibfnamefont
  {L.}~\bibnamefont {Morrison}}, \bibinfo {author} {\bibfnamefont
  {H.}~\bibnamefont {Stramel}}, \bibinfo {author} {\bibfnamefont
  {S.}~\bibnamefont {Yunus}}, \bibinfo {author} {\bibfnamefont
  {U.}~\bibnamefont {Zastrau}}, \bibinfo {author} {\bibfnamefont {S.~H.}\
  \bibnamefont {Glenzer}}, \bibinfo {author} {\bibfnamefont {G.}~\bibnamefont
  {Gregori}}, \bibinfo {author} {\bibfnamefont {D.~O.}\ \bibnamefont
  {Gericke}},\ and\ \bibinfo {author} {\bibfnamefont {B.}~\bibnamefont
  {Nagler}},\ }\href {https://doi.org/10.1038/s41586-025-09253-y} {\bibfield
  {journal} {\bibinfo  {journal} {Nature}\ }\textbf {\bibinfo {volume} {643}},\
  \bibinfo {pages} {950} (\bibinfo {year} {2025})}\BibitemShut {NoStop}%
\bibitem [{\citenamefont {Dorchies}\ \emph {et~al.}(2015)\citenamefont
  {Dorchies}, \citenamefont {Recoules}, \citenamefont {Bouchet}, \citenamefont
  {Fourment}, \citenamefont {Leguay}, \citenamefont {Cho}, \citenamefont
  {Engelhorn}, \citenamefont {Nakatsutsumi}, \citenamefont {Ozkan},
  \citenamefont {Tschentscher}, \citenamefont {Harmand}, \citenamefont
  {Toleikis}, \citenamefont {St\"ormer}, \citenamefont {Galtier}, \citenamefont
  {Lee}, \citenamefont {Nagler}, \citenamefont {Heimann},\ and\ \citenamefont
  {Gaudin}}]{Dorchies15}%
  \BibitemOpen
  \bibfield  {author} {\bibinfo {author} {\bibfnamefont {F.}~\bibnamefont
  {Dorchies}}, \bibinfo {author} {\bibfnamefont {V.}~\bibnamefont {Recoules}},
  \bibinfo {author} {\bibfnamefont {J.}~\bibnamefont {Bouchet}}, \bibinfo
  {author} {\bibfnamefont {C.}~\bibnamefont {Fourment}}, \bibinfo {author}
  {\bibfnamefont {P.~M.}\ \bibnamefont {Leguay}}, \bibinfo {author}
  {\bibfnamefont {B.~I.}\ \bibnamefont {Cho}}, \bibinfo {author} {\bibfnamefont
  {K.}~\bibnamefont {Engelhorn}}, \bibinfo {author} {\bibfnamefont
  {M.}~\bibnamefont {Nakatsutsumi}}, \bibinfo {author} {\bibfnamefont
  {C.}~\bibnamefont {Ozkan}}, \bibinfo {author} {\bibfnamefont
  {T.}~\bibnamefont {Tschentscher}}, \bibinfo {author} {\bibfnamefont
  {M.}~\bibnamefont {Harmand}}, \bibinfo {author} {\bibfnamefont
  {S.}~\bibnamefont {Toleikis}}, \bibinfo {author} {\bibfnamefont
  {M.}~\bibnamefont {St\"ormer}}, \bibinfo {author} {\bibfnamefont
  {E.}~\bibnamefont {Galtier}}, \bibinfo {author} {\bibfnamefont {H.~J.}\
  \bibnamefont {Lee}}, \bibinfo {author} {\bibfnamefont {B.}~\bibnamefont
  {Nagler}}, \bibinfo {author} {\bibfnamefont {P.~A.}\ \bibnamefont
  {Heimann}},\ and\ \bibinfo {author} {\bibfnamefont {J.}~\bibnamefont
  {Gaudin}},\ }\href {https://doi.org/10.1103/PhysRevB.92.144201} {\bibfield
  {journal} {\bibinfo  {journal} {Phys. Rev. B}\ }\textbf {\bibinfo {volume}
  {92}},\ \bibinfo {pages} {144201} (\bibinfo {year} {2015})}\BibitemShut
  {NoStop}%
\bibitem [{\citenamefont {Preston}\ \emph {et~al.}(2017)\citenamefont
  {Preston}, \citenamefont {Vinko}, \citenamefont {Ciricosta}, \citenamefont
  {Hollebon}, \citenamefont {Chung}, \citenamefont {Dakovski}, \citenamefont
  {Krzywinski}, \citenamefont {Minitti}, \citenamefont {Burian}, \citenamefont
  {Chalupsk\'y}, \citenamefont {H\'ajkov\'a}, \citenamefont {Juha},
  \citenamefont {Vozda}, \citenamefont {Zastrau}, \citenamefont {Lee},\ and\
  \citenamefont {Wark}}]{Preston17}%
  \BibitemOpen
  \bibfield  {author} {\bibinfo {author} {\bibfnamefont {T.~R.}\ \bibnamefont
  {Preston}}, \bibinfo {author} {\bibfnamefont {S.~M.}\ \bibnamefont {Vinko}},
  \bibinfo {author} {\bibfnamefont {O.}~\bibnamefont {Ciricosta}}, \bibinfo
  {author} {\bibfnamefont {P.}~\bibnamefont {Hollebon}}, \bibinfo {author}
  {\bibfnamefont {H.-K.}\ \bibnamefont {Chung}}, \bibinfo {author}
  {\bibfnamefont {G.~L.}\ \bibnamefont {Dakovski}}, \bibinfo {author}
  {\bibfnamefont {J.}~\bibnamefont {Krzywinski}}, \bibinfo {author}
  {\bibfnamefont {M.}~\bibnamefont {Minitti}}, \bibinfo {author} {\bibfnamefont
  {T.}~\bibnamefont {Burian}}, \bibinfo {author} {\bibfnamefont
  {J.}~\bibnamefont {Chalupsk\'y}}, \bibinfo {author} {\bibfnamefont
  {V.}~\bibnamefont {H\'ajkov\'a}}, \bibinfo {author} {\bibfnamefont
  {L.}~\bibnamefont {Juha}}, \bibinfo {author} {\bibfnamefont {V.}~\bibnamefont
  {Vozda}}, \bibinfo {author} {\bibfnamefont {U.}~\bibnamefont {Zastrau}},
  \bibinfo {author} {\bibfnamefont {R.~W.}\ \bibnamefont {Lee}},\ and\ \bibinfo
  {author} {\bibfnamefont {J.~S.}\ \bibnamefont {Wark}},\ }\href
  {https://doi.org/10.1103/PhysRevLett.119.085001} {\bibfield  {journal}
  {\bibinfo  {journal} {Phys. Rev. Lett.}\ }\textbf {\bibinfo {volume} {119}},\
  \bibinfo {pages} {085001} (\bibinfo {year} {2017})}\BibitemShut {NoStop}%
\bibitem [{\citenamefont {Mahieu}\ \emph {et~al.}(2018)\citenamefont {Mahieu},
  \citenamefont {Jourdain}, \citenamefont {Ta~Phuoc}, \citenamefont {Dorchies},
  \citenamefont {Goddet}, \citenamefont {Lifschitz}, \citenamefont {Renaudin},\
  and\ \citenamefont {Lecherbourg}}]{Mahieu2018}%
  \BibitemOpen
  \bibfield  {author} {\bibinfo {author} {\bibfnamefont {B.}~\bibnamefont
  {Mahieu}}, \bibinfo {author} {\bibfnamefont {N.}~\bibnamefont {Jourdain}},
  \bibinfo {author} {\bibfnamefont {K.}~\bibnamefont {Ta~Phuoc}}, \bibinfo
  {author} {\bibfnamefont {F.}~\bibnamefont {Dorchies}}, \bibinfo {author}
  {\bibfnamefont {J.-P.}\ \bibnamefont {Goddet}}, \bibinfo {author}
  {\bibfnamefont {A.}~\bibnamefont {Lifschitz}}, \bibinfo {author}
  {\bibfnamefont {P.}~\bibnamefont {Renaudin}},\ and\ \bibinfo {author}
  {\bibfnamefont {L.}~\bibnamefont {Lecherbourg}},\ }\href
  {https://doi.org/10.1038/s41467-018-05791-4} {\bibfield  {journal} {\bibinfo
  {journal} {Nature Communications}\ }\textbf {\bibinfo {volume} {9}},\
  \bibinfo {pages} {3276} (\bibinfo {year} {2018})}\BibitemShut {NoStop}%
\bibitem [{\citenamefont {Frydrych}\ \emph {et~al.}(2020)\citenamefont
  {Frydrych}, \citenamefont {Vorberger}, \citenamefont {Hartley}, \citenamefont
  {Schuster}, \citenamefont {Ramakrishna}, \citenamefont {Saunders},
  \citenamefont {van Driel}, \citenamefont {Falcone}, \citenamefont {Fletcher},
  \citenamefont {Galtier}, \citenamefont {Gamboa}, \citenamefont {Glenzer},
  \citenamefont {Granados}, \citenamefont {MacDonald}, \citenamefont
  {MacKinnon}, \citenamefont {McBride}, \citenamefont {Nam}, \citenamefont
  {Neumayer}, \citenamefont {Pak}, \citenamefont {Voigt}, \citenamefont {Roth},
  \citenamefont {Sun}, \citenamefont {Gericke}, \citenamefont {D{\"o}ppner},\
  and\ \citenamefont {Kraus}}]{Frydrych2020}%
  \BibitemOpen
  \bibfield  {author} {\bibinfo {author} {\bibfnamefont {S.}~\bibnamefont
  {Frydrych}}, \bibinfo {author} {\bibfnamefont {J.}~\bibnamefont {Vorberger}},
  \bibinfo {author} {\bibfnamefont {N.~J.}\ \bibnamefont {Hartley}}, \bibinfo
  {author} {\bibfnamefont {A.~K.}\ \bibnamefont {Schuster}}, \bibinfo {author}
  {\bibfnamefont {K.}~\bibnamefont {Ramakrishna}}, \bibinfo {author}
  {\bibfnamefont {A.~M.}\ \bibnamefont {Saunders}}, \bibinfo {author}
  {\bibfnamefont {T.}~\bibnamefont {van Driel}}, \bibinfo {author}
  {\bibfnamefont {R.~W.}\ \bibnamefont {Falcone}}, \bibinfo {author}
  {\bibfnamefont {L.~B.}\ \bibnamefont {Fletcher}}, \bibinfo {author}
  {\bibfnamefont {E.}~\bibnamefont {Galtier}}, \bibinfo {author} {\bibfnamefont
  {E.~J.}\ \bibnamefont {Gamboa}}, \bibinfo {author} {\bibfnamefont {S.~H.}\
  \bibnamefont {Glenzer}}, \bibinfo {author} {\bibfnamefont {E.}~\bibnamefont
  {Granados}}, \bibinfo {author} {\bibfnamefont {M.~J.}\ \bibnamefont
  {MacDonald}}, \bibinfo {author} {\bibfnamefont {A.~J.}\ \bibnamefont
  {MacKinnon}}, \bibinfo {author} {\bibfnamefont {E.~E.}\ \bibnamefont
  {McBride}}, \bibinfo {author} {\bibfnamefont {I.}~\bibnamefont {Nam}},
  \bibinfo {author} {\bibfnamefont {P.}~\bibnamefont {Neumayer}}, \bibinfo
  {author} {\bibfnamefont {A.}~\bibnamefont {Pak}}, \bibinfo {author}
  {\bibfnamefont {K.}~\bibnamefont {Voigt}}, \bibinfo {author} {\bibfnamefont
  {M.}~\bibnamefont {Roth}}, \bibinfo {author} {\bibfnamefont {P.}~\bibnamefont
  {Sun}}, \bibinfo {author} {\bibfnamefont {D.~O.}\ \bibnamefont {Gericke}},
  \bibinfo {author} {\bibfnamefont {T.}~\bibnamefont {D{\"o}ppner}},\ and\
  \bibinfo {author} {\bibfnamefont {D.}~\bibnamefont {Kraus}},\ }\href
  {https://doi.org/10.1038/s41467-020-16426-y} {\bibfield  {journal} {\bibinfo
  {journal} {Nature Communications}\ }\textbf {\bibinfo {volume} {11}},\
  \bibinfo {pages} {2620} (\bibinfo {year} {2020})}\BibitemShut {NoStop}%
\bibitem [{\citenamefont {Chen}\ \emph {et~al.}(2021)\citenamefont {Chen},
  \citenamefont {Na}, \citenamefont {Curry}, \citenamefont {Liang},
  \citenamefont {French}, \citenamefont {Descamps}, \citenamefont {DePonte},
  \citenamefont {Koralek}, \citenamefont {Kim}, \citenamefont {Lebovitz},
  \citenamefont {Nakatsutsumi}, \citenamefont {Ofori-Okai}, \citenamefont
  {Redmer}, \citenamefont {Roedel}, \citenamefont {Schörner}, \citenamefont
  {Skruszewicz}, \citenamefont {Sperling}, \citenamefont {Toleikis},
  \citenamefont {Mo},\ and\ \citenamefont {Glenzer}}]{Chen21}%
  \BibitemOpen
  \bibfield  {author} {\bibinfo {author} {\bibfnamefont {Z.}~\bibnamefont
  {Chen}}, \bibinfo {author} {\bibfnamefont {X.}~\bibnamefont {Na}}, \bibinfo
  {author} {\bibfnamefont {C.~B.}\ \bibnamefont {Curry}}, \bibinfo {author}
  {\bibfnamefont {S.}~\bibnamefont {Liang}}, \bibinfo {author} {\bibfnamefont
  {M.}~\bibnamefont {French}}, \bibinfo {author} {\bibfnamefont
  {A.}~\bibnamefont {Descamps}}, \bibinfo {author} {\bibfnamefont {D.~P.}\
  \bibnamefont {DePonte}}, \bibinfo {author} {\bibfnamefont {J.~D.}\
  \bibnamefont {Koralek}}, \bibinfo {author} {\bibfnamefont {J.~B.}\
  \bibnamefont {Kim}}, \bibinfo {author} {\bibfnamefont {S.}~\bibnamefont
  {Lebovitz}}, \bibinfo {author} {\bibfnamefont {M.}~\bibnamefont
  {Nakatsutsumi}}, \bibinfo {author} {\bibfnamefont {B.~K.}\ \bibnamefont
  {Ofori-Okai}}, \bibinfo {author} {\bibfnamefont {R.}~\bibnamefont {Redmer}},
  \bibinfo {author} {\bibfnamefont {C.}~\bibnamefont {Roedel}}, \bibinfo
  {author} {\bibfnamefont {M.}~\bibnamefont {Schörner}}, \bibinfo {author}
  {\bibfnamefont {S.}~\bibnamefont {Skruszewicz}}, \bibinfo {author}
  {\bibfnamefont {P.}~\bibnamefont {Sperling}}, \bibinfo {author}
  {\bibfnamefont {S.}~\bibnamefont {Toleikis}}, \bibinfo {author}
  {\bibfnamefont {M.~Z.}\ \bibnamefont {Mo}},\ and\ \bibinfo {author}
  {\bibfnamefont {S.~H.}\ \bibnamefont {Glenzer}},\ }\href
  {https://doi.org/10.1063/5.0043726} {\bibfield  {journal} {\bibinfo
  {journal} {Matter and Radiation at Extremes}\ }\textbf {\bibinfo {volume}
  {6}},\ \bibinfo {pages} {054401} (\bibinfo {year} {2021})}\BibitemShut
  {NoStop}%
\bibitem [{\citenamefont {D{\"o}ppner}\ \emph {et~al.}(2023)\citenamefont
  {D{\"o}ppner}, \citenamefont {Bethkenhagen}, \citenamefont {Kraus},
  \citenamefont {Neumayer}, \citenamefont {Chapman}, \citenamefont {Bachmann},
  \citenamefont {Baggott}, \citenamefont {B{\"o}hme}, \citenamefont {Divol},
  \citenamefont {Falcone}, \citenamefont {Fletcher}, \citenamefont {Landen},
  \citenamefont {MacDonald}, \citenamefont {Saunders}, \citenamefont
  {Sch{\"o}rner}, \citenamefont {Sterne}, \citenamefont {Vorberger},
  \citenamefont {Witte}, \citenamefont {Yi}, \citenamefont {Redmer},
  \citenamefont {Glenzer},\ and\ \citenamefont {Gericke}}]{Doppner2023}%
  \BibitemOpen
  \bibfield  {author} {\bibinfo {author} {\bibfnamefont {T.}~\bibnamefont
  {D{\"o}ppner}}, \bibinfo {author} {\bibfnamefont {M.}~\bibnamefont
  {Bethkenhagen}}, \bibinfo {author} {\bibfnamefont {D.}~\bibnamefont {Kraus}},
  \bibinfo {author} {\bibfnamefont {P.}~\bibnamefont {Neumayer}}, \bibinfo
  {author} {\bibfnamefont {D.~A.}\ \bibnamefont {Chapman}}, \bibinfo {author}
  {\bibfnamefont {B.}~\bibnamefont {Bachmann}}, \bibinfo {author}
  {\bibfnamefont {R.~A.}\ \bibnamefont {Baggott}}, \bibinfo {author}
  {\bibfnamefont {M.~P.}\ \bibnamefont {B{\"o}hme}}, \bibinfo {author}
  {\bibfnamefont {L.}~\bibnamefont {Divol}}, \bibinfo {author} {\bibfnamefont
  {R.~W.}\ \bibnamefont {Falcone}}, \bibinfo {author} {\bibfnamefont {L.~B.}\
  \bibnamefont {Fletcher}}, \bibinfo {author} {\bibfnamefont {O.~L.}\
  \bibnamefont {Landen}}, \bibinfo {author} {\bibfnamefont {M.~J.}\
  \bibnamefont {MacDonald}}, \bibinfo {author} {\bibfnamefont {A.~M.}\
  \bibnamefont {Saunders}}, \bibinfo {author} {\bibfnamefont {M.}~\bibnamefont
  {Sch{\"o}rner}}, \bibinfo {author} {\bibfnamefont {P.~A.}\ \bibnamefont
  {Sterne}}, \bibinfo {author} {\bibfnamefont {J.}~\bibnamefont {Vorberger}},
  \bibinfo {author} {\bibfnamefont {B.~B.~L.}\ \bibnamefont {Witte}}, \bibinfo
  {author} {\bibfnamefont {A.}~\bibnamefont {Yi}}, \bibinfo {author}
  {\bibfnamefont {R.}~\bibnamefont {Redmer}}, \bibinfo {author} {\bibfnamefont
  {S.~H.}\ \bibnamefont {Glenzer}},\ and\ \bibinfo {author} {\bibfnamefont
  {D.~O.}\ \bibnamefont {Gericke}},\ }\href
  {https://doi.org/10.1038/s41586-023-05996-8} {\bibfield  {journal} {\bibinfo
  {journal} {Nature}\ }\textbf {\bibinfo {volume} {618}},\ \bibinfo {pages}
  {270} (\bibinfo {year} {2023})}\BibitemShut {NoStop}%
\bibitem [{\citenamefont {Mercadier}\ \emph {et~al.}(2024)\citenamefont
  {Mercadier}, \citenamefont {Benediktovitch}, \citenamefont
  {Kru{\v{s}}i{\v{c}}}, \citenamefont {Kas}, \citenamefont {Schlappa},
  \citenamefont {Ag{\aa}ker}, \citenamefont {Carley}, \citenamefont {Fazio},
  \citenamefont {Gerasimova}, \citenamefont {Kim}, \citenamefont {Le~Guyader},
  \citenamefont {Mercurio}, \citenamefont {Parchenko}, \citenamefont {Rehr},
  \citenamefont {Rubensson}, \citenamefont {Serkez}, \citenamefont {Stransky},
  \citenamefont {Teichmann}, \citenamefont {Yin}, \citenamefont {{\v{Z}}itnik},
  \citenamefont {Scherz}, \citenamefont {Ziaja},\ and\ \citenamefont
  {Rohringer}}]{Mercadier2024}%
  \BibitemOpen
  \bibfield  {author} {\bibinfo {author} {\bibfnamefont {L.}~\bibnamefont
  {Mercadier}}, \bibinfo {author} {\bibfnamefont {A.}~\bibnamefont
  {Benediktovitch}}, \bibinfo {author} {\bibfnamefont {{\v{S}}.}~\bibnamefont
  {Kru{\v{s}}i{\v{c}}}}, \bibinfo {author} {\bibfnamefont {J.~J.}\ \bibnamefont
  {Kas}}, \bibinfo {author} {\bibfnamefont {J.}~\bibnamefont {Schlappa}},
  \bibinfo {author} {\bibfnamefont {M.}~\bibnamefont {Ag{\aa}ker}}, \bibinfo
  {author} {\bibfnamefont {R.}~\bibnamefont {Carley}}, \bibinfo {author}
  {\bibfnamefont {G.}~\bibnamefont {Fazio}}, \bibinfo {author} {\bibfnamefont
  {N.}~\bibnamefont {Gerasimova}}, \bibinfo {author} {\bibfnamefont {Y.~Y.}\
  \bibnamefont {Kim}}, \bibinfo {author} {\bibfnamefont {L.}~\bibnamefont
  {Le~Guyader}}, \bibinfo {author} {\bibfnamefont {G.}~\bibnamefont
  {Mercurio}}, \bibinfo {author} {\bibfnamefont {S.}~\bibnamefont {Parchenko}},
  \bibinfo {author} {\bibfnamefont {J.~J.}\ \bibnamefont {Rehr}}, \bibinfo
  {author} {\bibfnamefont {J.-E.}\ \bibnamefont {Rubensson}}, \bibinfo {author}
  {\bibfnamefont {S.}~\bibnamefont {Serkez}}, \bibinfo {author} {\bibfnamefont
  {M.}~\bibnamefont {Stransky}}, \bibinfo {author} {\bibfnamefont
  {M.}~\bibnamefont {Teichmann}}, \bibinfo {author} {\bibfnamefont
  {Z.}~\bibnamefont {Yin}}, \bibinfo {author} {\bibfnamefont {M.}~\bibnamefont
  {{\v{Z}}itnik}}, \bibinfo {author} {\bibfnamefont {A.}~\bibnamefont
  {Scherz}}, \bibinfo {author} {\bibfnamefont {B.}~\bibnamefont {Ziaja}},\ and\
  \bibinfo {author} {\bibfnamefont {N.}~\bibnamefont {Rohringer}},\ }\href
  {https://doi.org/10.1038/s41567-024-02587-w} {\bibfield  {journal} {\bibinfo
  {journal} {Nature Physics}\ }\textbf {\bibinfo {volume} {20}},\ \bibinfo
  {pages} {1564} (\bibinfo {year} {2024})}\BibitemShut {NoStop}%
\bibitem [{\citenamefont {Bespalov}\ \emph {et~al.}(2025)\citenamefont
  {Bespalov}, \citenamefont {Zastrau}, \citenamefont {Moldabekov},
  \citenamefont {Gawne}, \citenamefont {Dornheim}, \citenamefont {Meshhal},
  \citenamefont {Amouretti}, \citenamefont {Andrzejewski}, \citenamefont
  {Appel}, \citenamefont {Baehtz}, \citenamefont {Brambrink}, \citenamefont
  {Buakor}, \citenamefont {Camarda}, \citenamefont {Chin}, \citenamefont
  {Collins}, \citenamefont {Crepisson}, \citenamefont {Descamps}, \citenamefont
  {Eggert}, \citenamefont {Fletcher}, \citenamefont {Forte}, \citenamefont
  {Gregori}, \citenamefont {Harmand}, \citenamefont {Humphries}, \citenamefont
  {Hoeppner}, \citenamefont {Kuhlke}, \citenamefont {Lynn}, \citenamefont
  {Luetgert}, \citenamefont {Masruri}, \citenamefont {McBride}, \citenamefont
  {McWilliams}, \citenamefont {Mora}, \citenamefont {Naedler}, \citenamefont
  {Neumayer}, \citenamefont {Palmer}, \citenamefont {Pelka}, \citenamefont
  {Pennacchioni}, \citenamefont {Polsin}, \citenamefont {Prestwood},
  \citenamefont {Pukhareva}, \citenamefont {Qu}, \citenamefont {Ranjan},
  \citenamefont {Redmer}, \citenamefont {Roeper}, \citenamefont {Sahle},
  \citenamefont {Schumacher}, \citenamefont {Schwinkendorf}, \citenamefont
  {Sieber}, \citenamefont {Singleton}, \citenamefont {Smith}, \citenamefont
  {Sternemann}, \citenamefont {Stevens}, \citenamefont {Stevenson},
  \citenamefont {Strohm}, \citenamefont {Tang}, \citenamefont {Toncian},
  \citenamefont {Toncian}, \citenamefont {Tschentscher}, \citenamefont {Vinko},
  \citenamefont {Wark}, \citenamefont {Wilke}, \citenamefont {Kraus},\ and\
  \citenamefont {Preston}}]{Bespalov25}%
  \BibitemOpen
  \bibfield  {author} {\bibinfo {author} {\bibfnamefont {D.~S.}\ \bibnamefont
  {Bespalov}}, \bibinfo {author} {\bibfnamefont {U.}~\bibnamefont {Zastrau}},
  \bibinfo {author} {\bibfnamefont {Z.~A.}\ \bibnamefont {Moldabekov}},
  \bibinfo {author} {\bibfnamefont {T.}~\bibnamefont {Gawne}}, \bibinfo
  {author} {\bibfnamefont {T.}~\bibnamefont {Dornheim}}, \bibinfo {author}
  {\bibfnamefont {M.}~\bibnamefont {Meshhal}}, \bibinfo {author} {\bibfnamefont
  {A.}~\bibnamefont {Amouretti}}, \bibinfo {author} {\bibfnamefont
  {M.}~\bibnamefont {Andrzejewski}}, \bibinfo {author} {\bibfnamefont
  {K.}~\bibnamefont {Appel}}, \bibinfo {author} {\bibfnamefont
  {C.}~\bibnamefont {Baehtz}}, \bibinfo {author} {\bibfnamefont
  {E.}~\bibnamefont {Brambrink}}, \bibinfo {author} {\bibfnamefont
  {K.}~\bibnamefont {Buakor}}, \bibinfo {author} {\bibfnamefont
  {C.}~\bibnamefont {Camarda}}, \bibinfo {author} {\bibfnamefont
  {D.}~\bibnamefont {Chin}}, \bibinfo {author} {\bibfnamefont {G.}~\bibnamefont
  {Collins}}, \bibinfo {author} {\bibfnamefont {C.}~\bibnamefont {Crepisson}},
  \bibinfo {author} {\bibfnamefont {A.}~\bibnamefont {Descamps}}, \bibinfo
  {author} {\bibfnamefont {J.}~\bibnamefont {Eggert}}, \bibinfo {author}
  {\bibfnamefont {L.}~\bibnamefont {Fletcher}}, \bibinfo {author}
  {\bibfnamefont {A.}~\bibnamefont {Forte}}, \bibinfo {author} {\bibfnamefont
  {G.}~\bibnamefont {Gregori}}, \bibinfo {author} {\bibfnamefont
  {M.}~\bibnamefont {Harmand}}, \bibinfo {author} {\bibfnamefont {O.~S.}\
  \bibnamefont {Humphries}}, \bibinfo {author} {\bibfnamefont {H.}~\bibnamefont
  {Hoeppner}}, \bibinfo {author} {\bibfnamefont {J.}~\bibnamefont {Kuhlke}},
  \bibinfo {author} {\bibfnamefont {W.}~\bibnamefont {Lynn}}, \bibinfo {author}
  {\bibfnamefont {J.}~\bibnamefont {Luetgert}}, \bibinfo {author}
  {\bibfnamefont {M.}~\bibnamefont {Masruri}}, \bibinfo {author} {\bibfnamefont
  {E.~M.}\ \bibnamefont {McBride}}, \bibinfo {author} {\bibfnamefont {R.~S.}\
  \bibnamefont {McWilliams}}, \bibinfo {author} {\bibfnamefont {A.~A.~S.}\
  \bibnamefont {Mora}}, \bibinfo {author} {\bibfnamefont {J.-P.}\ \bibnamefont
  {Naedler}}, \bibinfo {author} {\bibfnamefont {P.}~\bibnamefont {Neumayer}},
  \bibinfo {author} {\bibfnamefont {C.}~\bibnamefont {Palmer}}, \bibinfo
  {author} {\bibfnamefont {A.}~\bibnamefont {Pelka}}, \bibinfo {author}
  {\bibfnamefont {L.}~\bibnamefont {Pennacchioni}}, \bibinfo {author}
  {\bibfnamefont {D.}~\bibnamefont {Polsin}}, \bibinfo {author} {\bibfnamefont
  {C.}~\bibnamefont {Prestwood}}, \bibinfo {author} {\bibfnamefont {N.~A.}\
  \bibnamefont {Pukhareva}}, \bibinfo {author} {\bibfnamefont {C.}~\bibnamefont
  {Qu}}, \bibinfo {author} {\bibfnamefont {D.}~\bibnamefont {Ranjan}}, \bibinfo
  {author} {\bibfnamefont {R.}~\bibnamefont {Redmer}}, \bibinfo {author}
  {\bibfnamefont {M.}~\bibnamefont {Roeper}}, \bibinfo {author} {\bibfnamefont
  {C.}~\bibnamefont {Sahle}}, \bibinfo {author} {\bibfnamefont
  {S.}~\bibnamefont {Schumacher}}, \bibinfo {author} {\bibfnamefont {J.-P.}\
  \bibnamefont {Schwinkendorf}}, \bibinfo {author} {\bibfnamefont {M.~J.}\
  \bibnamefont {Sieber}}, \bibinfo {author} {\bibfnamefont {M.}~\bibnamefont
  {Singleton}}, \bibinfo {author} {\bibfnamefont {E.}~\bibnamefont {Smith}},
  \bibinfo {author} {\bibfnamefont {C.}~\bibnamefont {Sternemann}}, \bibinfo
  {author} {\bibfnamefont {T.}~\bibnamefont {Stevens}}, \bibinfo {author}
  {\bibfnamefont {M.}~\bibnamefont {Stevenson}}, \bibinfo {author}
  {\bibfnamefont {C.}~\bibnamefont {Strohm}}, \bibinfo {author} {\bibfnamefont
  {M.}~\bibnamefont {Tang}}, \bibinfo {author} {\bibfnamefont {M.}~\bibnamefont
  {Toncian}}, \bibinfo {author} {\bibfnamefont {T.}~\bibnamefont {Toncian}},
  \bibinfo {author} {\bibfnamefont {T.}~\bibnamefont {Tschentscher}}, \bibinfo
  {author} {\bibfnamefont {S.}~\bibnamefont {Vinko}}, \bibinfo {author}
  {\bibfnamefont {J.}~\bibnamefont {Wark}}, \bibinfo {author} {\bibfnamefont
  {M.}~\bibnamefont {Wilke}}, \bibinfo {author} {\bibfnamefont
  {D.}~\bibnamefont {Kraus}},\ and\ \bibinfo {author} {\bibfnamefont {T.~R.}\
  \bibnamefont {Preston}},\ }\href {https://arxiv.org/abs/2509.10107} {\bibinfo
  {title} {Experimental validation of electron correlation models in warm dense
  matter}} (\bibinfo {year} {2025}),\ \Eprint
  {https://arxiv.org/abs/2509.10107} {arXiv:2509.10107 [physics.plasm-ph]}
  \BibitemShut {NoStop}%
\bibitem [{\citenamefont {Cordova}\ \emph {et~al.}(2025)\citenamefont
  {Cordova}, \citenamefont {Marley}, \citenamefont {Chin}, \citenamefont
  {London}, \citenamefont {Scott}, \citenamefont {Döppner}, \citenamefont
  {Beg}, \citenamefont {Coppari}, \citenamefont {Millot}, \citenamefont {Emig},
  \citenamefont {Hansen}, \citenamefont {Nilson}, \citenamefont {Sterne},\ and\
  \citenamefont {MacDonald}}]{Cordova25}%
  \BibitemOpen
  \bibfield  {author} {\bibinfo {author} {\bibfnamefont {T.}~\bibnamefont
  {Cordova}}, \bibinfo {author} {\bibfnamefont {E.~V.}\ \bibnamefont {Marley}},
  \bibinfo {author} {\bibfnamefont {D.~A.}\ \bibnamefont {Chin}}, \bibinfo
  {author} {\bibfnamefont {R.~A.}\ \bibnamefont {London}}, \bibinfo {author}
  {\bibfnamefont {H.~A.}\ \bibnamefont {Scott}}, \bibinfo {author}
  {\bibfnamefont {T.}~\bibnamefont {Döppner}}, \bibinfo {author}
  {\bibfnamefont {F.~N.}\ \bibnamefont {Beg}}, \bibinfo {author} {\bibfnamefont
  {F.}~\bibnamefont {Coppari}}, \bibinfo {author} {\bibfnamefont
  {M.}~\bibnamefont {Millot}}, \bibinfo {author} {\bibfnamefont
  {J.}~\bibnamefont {Emig}}, \bibinfo {author} {\bibfnamefont {S.~B.}\
  \bibnamefont {Hansen}}, \bibinfo {author} {\bibfnamefont {P.~M.}\
  \bibnamefont {Nilson}}, \bibinfo {author} {\bibfnamefont {P.}~\bibnamefont
  {Sterne}},\ and\ \bibinfo {author} {\bibfnamefont {M.~J.}\ \bibnamefont
  {MacDonald}},\ }\href {https://arxiv.org/abs/2509.13272} {\bibinfo {title}
  {Ionization and temperature measurements in warm dense copper using x-ray
  absorption spectroscopy}} (\bibinfo {year} {2025}),\ \Eprint
  {https://arxiv.org/abs/2509.13272} {arXiv:2509.13272 [hep-ex]} \BibitemShut
  {NoStop}%
\bibitem [{\citenamefont {Dufty}\ \emph {et~al.}(2018)\citenamefont {Dufty},
  \citenamefont {Wrighton}, \citenamefont {Luo},\ and\ \citenamefont
  {Trickey}}]{Dufty2018}%
  \BibitemOpen
  \bibfield  {author} {\bibinfo {author} {\bibfnamefont {J.}~\bibnamefont
  {Dufty}}, \bibinfo {author} {\bibfnamefont {J.}~\bibnamefont {Wrighton}},
  \bibinfo {author} {\bibfnamefont {K.}~\bibnamefont {Luo}},\ and\ \bibinfo
  {author} {\bibfnamefont {S.}~\bibnamefont {Trickey}},\ }\href
  {https://doi.org/https://doi.org/10.1002/ctpp.201700102} {\bibfield
  {journal} {\bibinfo  {journal} {Contributions to Plasma Physics}\ }\textbf
  {\bibinfo {volume} {58}},\ \bibinfo {pages} {150} (\bibinfo {year}
  {2018})}\BibitemShut {NoStop}%
\bibitem [{\citenamefont {White}\ \emph {et~al.}(2024)\citenamefont {White},
  \citenamefont {Craven}, \citenamefont {Sharma},\ and\ \citenamefont
  {Collins}}]{White2024}%
  \BibitemOpen
  \bibfield  {author} {\bibinfo {author} {\bibfnamefont {A.~J.}\ \bibnamefont
  {White}}, \bibinfo {author} {\bibfnamefont {G.~T.}\ \bibnamefont {Craven}},
  \bibinfo {author} {\bibfnamefont {V.}~\bibnamefont {Sharma}},\ and\ \bibinfo
  {author} {\bibfnamefont {L.~A.}\ \bibnamefont {Collins}},\ }\href
  {https://doi.org/10.1063/5.0198003} {\bibfield  {journal} {\bibinfo
  {journal} {Physics of Plasmas}\ }\textbf {\bibinfo {volume} {31}},\ \bibinfo
  {pages} {042706} (\bibinfo {year} {2024})}\BibitemShut {NoStop}%
\bibitem [{\citenamefont {Allen}(2006)}]{ALLEN2006165}%
  \BibitemOpen
  \bibfield  {author} {\bibinfo {author} {\bibfnamefont {P.}~\bibnamefont
  {Allen}},\ }in\ \href
  {https://doi.org/https://doi.org/10.1016/S1572-0934(06)02006-3} {\emph
  {\bibinfo {booktitle} {{Conceptual Foundations of Materials}}}},\ \bibinfo
  {series} {Contemporary Concepts of Condensed Matter Science}, Vol.~\bibinfo
  {volume} {2},\ \bibinfo {editor} {edited by\ \bibinfo {editor} {\bibfnamefont
  {S.~G.}\ \bibnamefont {Louie}}\ and\ \bibinfo {editor} {\bibfnamefont
  {M.~L.}\ \bibnamefont {Cohen}}}\ (\bibinfo  {publisher} {Elsevier},\ \bibinfo
  {year} {2006})\ pp.\ \bibinfo {pages} {165--218}\BibitemShut {NoStop}%
\bibitem [{\citenamefont {Kubo}(1957)}]{Kubo1957}%
  \BibitemOpen
  \bibfield  {author} {\bibinfo {author} {\bibfnamefont {R.}~\bibnamefont
  {Kubo}},\ }\href {https://doi.org/10.1143/JPSJ.12.570} {\bibfield  {journal}
  {\bibinfo  {journal} {Journal of the Physical Society of Japan}\ }\textbf
  {\bibinfo {volume} {12}},\ \bibinfo {pages} {570} (\bibinfo {year}
  {1957})}\BibitemShut {NoStop}%
\bibitem [{\citenamefont {Greenwood}(1958)}]{Greenwood1958}%
  \BibitemOpen
  \bibfield  {author} {\bibinfo {author} {\bibfnamefont {D.~A.}\ \bibnamefont
  {Greenwood}},\ }\href {https://doi.org/10.1088/0370-1328/71/4/306} {\bibfield
   {journal} {\bibinfo  {journal} {Proceedings of the Physical Society}\
  }\textbf {\bibinfo {volume} {71}},\ \bibinfo {pages} {585} (\bibinfo {year}
  {1958})}\BibitemShut {NoStop}%
\bibitem [{\citenamefont {White}\ and\ \citenamefont
  {Collins}(2020)}]{White2020}%
  \BibitemOpen
  \bibfield  {author} {\bibinfo {author} {\bibfnamefont {A.~J.}\ \bibnamefont
  {White}}\ and\ \bibinfo {author} {\bibfnamefont {L.~A.}\ \bibnamefont
  {Collins}},\ }\href {https://doi.org/10.1103/PhysRevLett.125.055002}
  {\bibfield  {journal} {\bibinfo  {journal} {Phys. Rev. Lett.}\ }\textbf
  {\bibinfo {volume} {125}},\ \bibinfo {pages} {055002} (\bibinfo {year}
  {2020})}\BibitemShut {NoStop}%
\bibitem [{\citenamefont {Sharma}\ \emph {et~al.}(2023)\citenamefont {Sharma},
  \citenamefont {Collins},\ and\ \citenamefont {White}}]{Sharma2023}%
  \BibitemOpen
  \bibfield  {author} {\bibinfo {author} {\bibfnamefont {V.}~\bibnamefont
  {Sharma}}, \bibinfo {author} {\bibfnamefont {L.~A.}\ \bibnamefont
  {Collins}},\ and\ \bibinfo {author} {\bibfnamefont {A.~J.}\ \bibnamefont
  {White}},\ }\href {https://doi.org/10.1103/PhysRevE.108.L023201} {\bibfield
  {journal} {\bibinfo  {journal} {Phys. Rev. E}\ }\textbf {\bibinfo {volume}
  {108}},\ \bibinfo {pages} {L023201} (\bibinfo {year} {2023})}\BibitemShut
  {NoStop}%
\bibitem [{\citenamefont {Sharma}\ and\ \citenamefont
  {White}(2025)}]{Sharma2025}%
  \BibitemOpen
  \bibfield  {author} {\bibinfo {author} {\bibfnamefont {V.}~\bibnamefont
  {Sharma}}\ and\ \bibinfo {author} {\bibfnamefont {A.~J.}\ \bibnamefont
  {White}},\ }\href {https://doi.org/10.1103/PhysRevLett.134.095102} {\bibfield
   {journal} {\bibinfo  {journal} {Phys. Rev. Lett.}\ }\textbf {\bibinfo
  {volume} {134}},\ \bibinfo {pages} {095102} (\bibinfo {year}
  {2025})}\BibitemShut {NoStop}%
\bibitem [{\citenamefont {Bl\"{o}chl}(1994)}]{Blochl1994}%
  \BibitemOpen
  \bibfield  {author} {\bibinfo {author} {\bibfnamefont {P.~E.}\ \bibnamefont
  {Bl\"{o}chl}},\ }\href {https://doi.org/10.1103/PhysRevB.50.17953} {\bibfield
   {journal} {\bibinfo  {journal} {Phys. Rev. B}\ }\textbf {\bibinfo {volume}
  {50}},\ \bibinfo {pages} {17953} (\bibinfo {year} {1994})}\BibitemShut
  {NoStop}%
\bibitem [{\citenamefont {Kresse}\ and\ \citenamefont
  {Joubert}(1999{\natexlab{a}})}]{Kresse1999}%
  \BibitemOpen
  \bibfield  {author} {\bibinfo {author} {\bibfnamefont {G.}~\bibnamefont
  {Kresse}}\ and\ \bibinfo {author} {\bibfnamefont {D.}~\bibnamefont
  {Joubert}},\ }\href {https://doi.org/10.1103/PhysRevB.59.1758} {\bibfield
  {journal} {\bibinfo  {journal} {Phys. Rev. B}\ }\textbf {\bibinfo {volume}
  {59}},\ \bibinfo {pages} {1758} (\bibinfo {year}
  {1999}{\natexlab{a}})}\BibitemShut {NoStop}%
\bibitem [{\citenamefont {Perdew}\ \emph {et~al.}(1996)\citenamefont {Perdew},
  \citenamefont {Burke},\ and\ \citenamefont {Ernzerhof}}]{PBE1996}%
  \BibitemOpen
  \bibfield  {author} {\bibinfo {author} {\bibfnamefont {J.~P.}\ \bibnamefont
  {Perdew}}, \bibinfo {author} {\bibfnamefont {K.}~\bibnamefont {Burke}},\ and\
  \bibinfo {author} {\bibfnamefont {M.}~\bibnamefont {Ernzerhof}},\ }\href
  {https://doi.org/10.1103/PhysRevLett.77.3865} {\bibfield  {journal} {\bibinfo
   {journal} {Phys. Rev. Lett.}\ }\textbf {\bibinfo {volume} {77}},\ \bibinfo
  {pages} {3865} (\bibinfo {year} {1996})}\BibitemShut {NoStop}%
\bibitem [{\citenamefont {White}(2021)}]{shred}%
  \BibitemOpen
  \bibfield  {author} {\bibinfo {author} {\bibfnamefont {A.~J.}\ \bibnamefont
  {White}},\ }\href {https://github.com/lanl/SHRED/} {\bibinfo {title} {{SHRED:
  Stochastic and Hybrid Representation Electronic structure by Density
  functional theory, a plane-wave DFT code employing Kohn-Sham, orbital-free,
  stochastic, and mixed stochastic-deterministic DFT methods.}}} (\bibinfo
  {year} {2021})\BibitemShut {NoStop}%
\bibitem [{\citenamefont {Minary}\ \emph {et~al.}(2003)\citenamefont {Minary},
  \citenamefont {Martyna},\ and\ \citenamefont {Tuckerman}}]{Minary03}%
  \BibitemOpen
  \bibfield  {author} {\bibinfo {author} {\bibfnamefont {P.}~\bibnamefont
  {Minary}}, \bibinfo {author} {\bibfnamefont {G.~J.}\ \bibnamefont
  {Martyna}},\ and\ \bibinfo {author} {\bibfnamefont {M.~E.}\ \bibnamefont
  {Tuckerman}},\ }\href {https://doi.org/10.1063/1.1534583} {\bibfield
  {journal} {\bibinfo  {journal} {The Journal of Chemical Physics}\ }\textbf
  {\bibinfo {volume} {118}},\ \bibinfo {pages} {2527} (\bibinfo {year}
  {2003})}\BibitemShut {NoStop}%
\bibitem [{\citenamefont {Holst}\ \emph
  {et~al.}(2011{\natexlab{a}})\citenamefont {Holst}, \citenamefont {French},\
  and\ \citenamefont {Redmer}}]{Holst2011}%
  \BibitemOpen
  \bibfield  {author} {\bibinfo {author} {\bibfnamefont {B.}~\bibnamefont
  {Holst}}, \bibinfo {author} {\bibfnamefont {M.}~\bibnamefont {French}},\ and\
  \bibinfo {author} {\bibfnamefont {R.}~\bibnamefont {Redmer}},\ }\href
  {https://doi.org/10.1103/PhysRevB.83.235120} {\bibfield  {journal} {\bibinfo
  {journal} {Phys. Rev. B}\ }\textbf {\bibinfo {volume} {83}},\ \bibinfo
  {pages} {235120} (\bibinfo {year} {2011}{\natexlab{a}})}\BibitemShut
  {NoStop}%
\bibitem [{\citenamefont {Holst}\ \emph
  {et~al.}(2011{\natexlab{b}})\citenamefont {Holst}, \citenamefont {French},\
  and\ \citenamefont {Redmer}}]{Holst11}%
  \BibitemOpen
  \bibfield  {author} {\bibinfo {author} {\bibfnamefont {B.}~\bibnamefont
  {Holst}}, \bibinfo {author} {\bibfnamefont {M.}~\bibnamefont {French}},\ and\
  \bibinfo {author} {\bibfnamefont {R.}~\bibnamefont {Redmer}},\ }\href
  {https://doi.org/10.1103/PhysRevB.83.235120} {\bibfield  {journal} {\bibinfo
  {journal} {Phys. Rev. B}\ }\textbf {\bibinfo {volume} {83}},\ \bibinfo
  {pages} {235120} (\bibinfo {year} {2011}{\natexlab{b}})}\BibitemShut
  {NoStop}%
\bibitem [{\citenamefont {Mermin}(1965)}]{Mermin1965}%
  \BibitemOpen
  \bibfield  {author} {\bibinfo {author} {\bibfnamefont {N.~D.}\ \bibnamefont
  {Mermin}},\ }\href {https://doi.org/10.1103/PhysRev.137.A1441} {\bibfield
  {journal} {\bibinfo  {journal} {Phys. Rev.}\ }\textbf {\bibinfo {volume}
  {137}},\ \bibinfo {pages} {A1441} (\bibinfo {year} {1965})}\BibitemShut
  {NoStop}%
\bibitem [{\citenamefont {Hutchinson}(1990)}]{Hutchinson1990}%
  \BibitemOpen
  \bibfield  {author} {\bibinfo {author} {\bibfnamefont {M.}~\bibnamefont
  {Hutchinson}},\ }\href {https://doi.org/10.1080/03610919008812866} {\bibfield
   {journal} {\bibinfo  {journal} {Communications in Statistics - Simulation
  and Computation}\ }\textbf {\bibinfo {volume} {19}},\ \bibinfo {pages} {433}
  (\bibinfo {year} {1990})}\BibitemShut {NoStop}%
\bibitem [{\citenamefont {Cytter}\ \emph {et~al.}(2019)\citenamefont {Cytter},
  \citenamefont {Rabani}, \citenamefont {Neuhauser}, \citenamefont {Preising},
  \citenamefont {Redmer},\ and\ \citenamefont {Baer}}]{Cytter2019}%
  \BibitemOpen
  \bibfield  {author} {\bibinfo {author} {\bibfnamefont {Y.}~\bibnamefont
  {Cytter}}, \bibinfo {author} {\bibfnamefont {E.}~\bibnamefont {Rabani}},
  \bibinfo {author} {\bibfnamefont {D.}~\bibnamefont {Neuhauser}}, \bibinfo
  {author} {\bibfnamefont {M.}~\bibnamefont {Preising}}, \bibinfo {author}
  {\bibfnamefont {R.}~\bibnamefont {Redmer}},\ and\ \bibinfo {author}
  {\bibfnamefont {R.}~\bibnamefont {Baer}},\ }\href
  {https://doi.org/10.1103/PhysRevB.100.195101} {\bibfield  {journal} {\bibinfo
   {journal} {Phys. Rev. B}\ }\textbf {\bibinfo {volume} {100}},\ \bibinfo
  {pages} {195101} (\bibinfo {year} {2019})}\BibitemShut {NoStop}%
\bibitem [{\citenamefont {Park}\ and\ \citenamefont {Light}(1986)}]{Park1986}%
  \BibitemOpen
  \bibfield  {author} {\bibinfo {author} {\bibfnamefont {T.~J.}\ \bibnamefont
  {Park}}\ and\ \bibinfo {author} {\bibfnamefont {J.~C.}\ \bibnamefont
  {Light}},\ }\href {https://doi.org/10.1063/1.451548} {\bibfield  {journal}
  {\bibinfo  {journal} {The Journal of Chemical Physics}\ }\textbf {\bibinfo
  {volume} {85}},\ \bibinfo {pages} {5870} (\bibinfo {year}
  {1986})}\BibitemShut {NoStop}%
\bibitem [{\citenamefont {Kosloff}(1988)}]{Kosloff1988}%
  \BibitemOpen
  \bibfield  {author} {\bibinfo {author} {\bibfnamefont {R.}~\bibnamefont
  {Kosloff}},\ }\href {https://doi.org/10.1021/j100319a003} {\bibfield
  {journal} {\bibinfo  {journal} {The Journal of Physical Chemistry}\ }\textbf
  {\bibinfo {volume} {92}},\ \bibinfo {pages} {2087} (\bibinfo {year}
  {1988})}\BibitemShut {NoStop}%
\bibitem [{\citenamefont {White}\ \emph {et~al.}(2022)\citenamefont {White},
  \citenamefont {Collins}, \citenamefont {Nichols},\ and\ \citenamefont
  {Hu}}]{White22}%
  \BibitemOpen
  \bibfield  {author} {\bibinfo {author} {\bibfnamefont {A.~J.}\ \bibnamefont
  {White}}, \bibinfo {author} {\bibfnamefont {L.~A.}\ \bibnamefont {Collins}},
  \bibinfo {author} {\bibfnamefont {K.}~\bibnamefont {Nichols}},\ and\ \bibinfo
  {author} {\bibfnamefont {S.~X.}\ \bibnamefont {Hu}},\ }\href
  {https://doi.org/10.1088/1361-648X/ac4f1a} {\bibfield  {journal} {\bibinfo
  {journal} {Journal of Physics: Condensed Matter}\ }\textbf {\bibinfo {volume}
  {34}},\ \bibinfo {pages} {174001} (\bibinfo {year} {2022})}\BibitemShut
  {NoStop}%
\bibitem [{\citenamefont {Baczewski}\ \emph {et~al.}(2016)\citenamefont
  {Baczewski}, \citenamefont {Shulenburger}, \citenamefont {Desjarlais},
  \citenamefont {Hansen},\ and\ \citenamefont {Magyar}}]{Baczewski16}%
  \BibitemOpen
  \bibfield  {author} {\bibinfo {author} {\bibfnamefont {A.~D.}\ \bibnamefont
  {Baczewski}}, \bibinfo {author} {\bibfnamefont {L.}~\bibnamefont
  {Shulenburger}}, \bibinfo {author} {\bibfnamefont {M.~P.}\ \bibnamefont
  {Desjarlais}}, \bibinfo {author} {\bibfnamefont {S.~B.}\ \bibnamefont
  {Hansen}},\ and\ \bibinfo {author} {\bibfnamefont {R.~J.}\ \bibnamefont
  {Magyar}},\ }\href {https://doi.org/10.1103/PhysRevLett.116.115004}
  {\bibfield  {journal} {\bibinfo  {journal} {Phys. Rev. Lett.}\ }\textbf
  {\bibinfo {volume} {116}},\ \bibinfo {pages} {115004} (\bibinfo {year}
  {2016})}\BibitemShut {NoStop}%
\bibitem [{\citenamefont {Kononov}\ \emph {et~al.}(2022)\citenamefont
  {Kononov}, \citenamefont {Lee}, \citenamefont {dos Santos}, \citenamefont
  {Robinson}, \citenamefont {Yao}, \citenamefont {Yao}, \citenamefont
  {Andrade}, \citenamefont {Baczewski}, \citenamefont {Constantinescu},
  \citenamefont {Correa}, \citenamefont {Kanai}, \citenamefont {Modine},\ and\
  \citenamefont {Schleife}}]{Kononov2022}%
  \BibitemOpen
  \bibfield  {author} {\bibinfo {author} {\bibfnamefont {A.}~\bibnamefont
  {Kononov}}, \bibinfo {author} {\bibfnamefont {C.-W.}\ \bibnamefont {Lee}},
  \bibinfo {author} {\bibfnamefont {T.~P.}\ \bibnamefont {dos Santos}},
  \bibinfo {author} {\bibfnamefont {B.}~\bibnamefont {Robinson}}, \bibinfo
  {author} {\bibfnamefont {Y.}~\bibnamefont {Yao}}, \bibinfo {author}
  {\bibfnamefont {Y.}~\bibnamefont {Yao}}, \bibinfo {author} {\bibfnamefont
  {X.}~\bibnamefont {Andrade}}, \bibinfo {author} {\bibfnamefont {A.~D.}\
  \bibnamefont {Baczewski}}, \bibinfo {author} {\bibfnamefont {E.}~\bibnamefont
  {Constantinescu}}, \bibinfo {author} {\bibfnamefont {A.~A.}\ \bibnamefont
  {Correa}}, \bibinfo {author} {\bibfnamefont {Y.}~\bibnamefont {Kanai}},
  \bibinfo {author} {\bibfnamefont {N.}~\bibnamefont {Modine}},\ and\ \bibinfo
  {author} {\bibfnamefont {A.}~\bibnamefont {Schleife}},\ }\href
  {https://doi.org/10.1557/s43579-022-00273-7} {\bibfield  {journal} {\bibinfo
  {journal} {MRS Communications}\ }\textbf {\bibinfo {volume} {12}},\ \bibinfo
  {pages} {1002} (\bibinfo {year} {2022})}\BibitemShut {NoStop}%
\bibitem [{\citenamefont {Andrade}\ \emph {et~al.}(2018)\citenamefont
  {Andrade}, \citenamefont {Hamel},\ and\ \citenamefont
  {Correa}}]{Andrade2018}%
  \BibitemOpen
  \bibfield  {author} {\bibinfo {author} {\bibfnamefont {X.}~\bibnamefont
  {Andrade}}, \bibinfo {author} {\bibfnamefont {S.}~\bibnamefont {Hamel}},\
  and\ \bibinfo {author} {\bibfnamefont {A.~A.}\ \bibnamefont {Correa}},\
  }\href {https://doi.org/10.1140/epjb/e2018-90291-5} {\bibfield  {journal}
  {\bibinfo  {journal} {The European Physical Journal B}\ }\textbf {\bibinfo
  {volume} {91}},\ \bibinfo {pages} {229} (\bibinfo {year} {2018})}\BibitemShut
  {NoStop}%
\bibitem [{\citenamefont {Ramakrishna}\ \emph {et~al.}(2023)\citenamefont
  {Ramakrishna}, \citenamefont {Lokamani}, \citenamefont {Baczewski},
  \citenamefont {Vorberger},\ and\ \citenamefont {Cangi}}]{Ramakrishna23}%
  \BibitemOpen
  \bibfield  {author} {\bibinfo {author} {\bibfnamefont {K.}~\bibnamefont
  {Ramakrishna}}, \bibinfo {author} {\bibfnamefont {M.}~\bibnamefont
  {Lokamani}}, \bibinfo {author} {\bibfnamefont {A.}~\bibnamefont {Baczewski}},
  \bibinfo {author} {\bibfnamefont {J.}~\bibnamefont {Vorberger}},\ and\
  \bibinfo {author} {\bibfnamefont {A.}~\bibnamefont {Cangi}},\ }\href
  {https://doi.org/10.1103/PhysRevB.107.115131} {\bibfield  {journal} {\bibinfo
   {journal} {Phys. Rev. B}\ }\textbf {\bibinfo {volume} {107}},\ \bibinfo
  {pages} {115131} (\bibinfo {year} {2023})}\BibitemShut {NoStop}%
\bibitem [{\citenamefont {Ramakrishna}\ \emph {et~al.}(2024)\citenamefont
  {Ramakrishna}, \citenamefont {Lokamani},\ and\ \citenamefont
  {Cangi}}]{Ramakrishna24}%
  \BibitemOpen
  \bibfield  {author} {\bibinfo {author} {\bibfnamefont {K.}~\bibnamefont
  {Ramakrishna}}, \bibinfo {author} {\bibfnamefont {M.}~\bibnamefont
  {Lokamani}},\ and\ \bibinfo {author} {\bibfnamefont {A.}~\bibnamefont
  {Cangi}},\ }\href {https://doi.org/10.1088/2516-1075/ad912b} {\bibfield
  {journal} {\bibinfo  {journal} {Electronic Structure}\ }\textbf {\bibinfo
  {volume} {6}},\ \bibinfo {pages} {045008} (\bibinfo {year}
  {2024})}\BibitemShut {NoStop}%
\bibitem [{\citenamefont {White}(2025)}]{White25}%
  \BibitemOpen
  \bibfield  {author} {\bibinfo {author} {\bibfnamefont {A.~J.}\ \bibnamefont
  {White}},\ }\href {https://doi.org/10.1088/2516-1075/adad24} {\bibfield
  {journal} {\bibinfo  {journal} {Electronic Structure}\ }\textbf {\bibinfo
  {volume} {7}},\ \bibinfo {pages} {014001} (\bibinfo {year}
  {2025})}\BibitemShut {NoStop}%
\bibitem [{\citenamefont {Bl\"ochl}(1994)}]{Blochl94}%
  \BibitemOpen
  \bibfield  {author} {\bibinfo {author} {\bibfnamefont {P.~E.}\ \bibnamefont
  {Bl\"ochl}},\ }\href {https://doi.org/10.1103/PhysRevB.50.17953} {\bibfield
  {journal} {\bibinfo  {journal} {Phys. Rev. B}\ }\textbf {\bibinfo {volume}
  {50}},\ \bibinfo {pages} {17953} (\bibinfo {year} {1994})}\BibitemShut
  {NoStop}%
\bibitem [{\citenamefont {Rostgaard}(2009)}]{Rostgaard09}%
  \BibitemOpen
  \bibfield  {author} {\bibinfo {author} {\bibfnamefont {C.}~\bibnamefont
  {Rostgaard}},\ }\href {https://arxiv.org/abs/0910.1921} {\bibinfo {title}
  {The projector augmented-wave method}} (\bibinfo {year} {2009}),\ \Eprint
  {https://arxiv.org/abs/0910.1921} {arXiv:0910.1921 [cond-mat.mtrl-sci]}
  \BibitemShut {NoStop}%
\bibitem [{\citenamefont {Kresse}\ and\ \citenamefont
  {Joubert}(1999{\natexlab{b}})}]{Kresse99}%
  \BibitemOpen
  \bibfield  {author} {\bibinfo {author} {\bibfnamefont {G.}~\bibnamefont
  {Kresse}}\ and\ \bibinfo {author} {\bibfnamefont {D.}~\bibnamefont
  {Joubert}},\ }\href {https://doi.org/10.1103/PhysRevB.59.1758} {\bibfield
  {journal} {\bibinfo  {journal} {Phys. Rev. B}\ }\textbf {\bibinfo {volume}
  {59}},\ \bibinfo {pages} {1758} (\bibinfo {year}
  {1999}{\natexlab{b}})}\BibitemShut {NoStop}%
\bibitem [{\citenamefont {Li}\ and\ \citenamefont {Neuhauser}(2020)}]{Li2020}%
  \BibitemOpen
  \bibfield  {author} {\bibinfo {author} {\bibfnamefont {W.}~\bibnamefont
  {Li}}\ and\ \bibinfo {author} {\bibfnamefont {D.}~\bibnamefont {Neuhauser}},\
  }\href {https://doi.org/10.1103/PhysRevB.102.195118} {\bibfield  {journal}
  {\bibinfo  {journal} {Phys. Rev. B}\ }\textbf {\bibinfo {volume} {102}},\
  \bibinfo {pages} {195118} (\bibinfo {year} {2020})}\BibitemShut {NoStop}%
\bibitem [{\citenamefont {Woodbury}(1949)}]{Woodbury:1949}%
  \BibitemOpen
  \bibfield  {author} {\bibinfo {author} {\bibfnamefont {M.~A.}\ \bibnamefont
  {Woodbury}},\ }\href@noop {} {\emph {\bibinfo {title} {The {S}tability of
  {O}ut-{I}nput {M}atrices}}}\ (\bibinfo  {publisher} {Chicago, Ill.},\
  \bibinfo {year} {1949})\ p.~\bibinfo {pages} {5}\BibitemShut {NoStop}%
\bibitem [{\citenamefont {Woodbury}(1950)}]{Woodbury:1950}%
  \BibitemOpen
  \bibfield  {author} {\bibinfo {author} {\bibfnamefont {M.~A.}\ \bibnamefont
  {Woodbury}},\ }\href@noop {} {\emph {\bibinfo {title} {Inverting modified
  matrices}}}\ (\bibinfo  {publisher} {Princeton University, Princeton, NJ},\
  \bibinfo {year} {1950})\ p.~\bibinfo {pages} {4},\ \bibinfo {note}
  {statistical Research Group, Memo. Rep. no. 42,}\BibitemShut {NoStop}%
\bibitem [{\citenamefont {Higham}(2002)}]{Higham:2002}%
  \BibitemOpen
  \bibfield  {author} {\bibinfo {author} {\bibfnamefont {N.~J.}\ \bibnamefont
  {Higham}},\ }\href {https://doi.org/10.1137/1.9780898718027} {\emph {\bibinfo
  {title} {Accuracy and stability of numerical algorithms}}},\ \bibinfo
  {edition} {2nd}\ ed.\ (\bibinfo  {publisher} {Society for Industrial and
  Applied Mathematics (SIAM), Philadelphia, PA},\ \bibinfo {year} {2002})\ pp.\
  \bibinfo {pages} {xxx+680}\BibitemShut {NoStop}%
\bibitem [{\citenamefont {Golub}\ and\ \citenamefont
  {Van~Loan}(2013)}]{Golub2013-dz}%
  \BibitemOpen
  \bibfield  {author} {\bibinfo {author} {\bibfnamefont {G.~H.}\ \bibnamefont
  {Golub}}\ and\ \bibinfo {author} {\bibfnamefont {C.~F.}\ \bibnamefont
  {Van~Loan}},\ }\href@noop {} {\emph {\bibinfo {title} {Matrix
  Computations}}},\ \bibinfo {edition} {4th}\ ed.,\ Johns Hopkins Studies in
  the Mathematical Sciences\ (\bibinfo  {publisher} {Johns Hopkins University
  Press},\ \bibinfo {address} {Baltimore, MD},\ \bibinfo {year}
  {2013})\BibitemShut {NoStop}%
\bibitem [{\citenamefont {Lign{\`e}res}\ and\ \citenamefont
  {Carter}(2005)}]{Ligneres2005}%
  \BibitemOpen
  \bibfield  {author} {\bibinfo {author} {\bibfnamefont {V.~L.}\ \bibnamefont
  {Lign{\`e}res}}\ and\ \bibinfo {author} {\bibfnamefont {E.~A.}\ \bibnamefont
  {Carter}},\ }\bibinfo {title} {An introduction to orbital-free density
  functional theory},\ in\ \href {https://doi.org/10.1007/978-1-4020-3286-8_9}
  {\emph {\bibinfo {booktitle} {Handbook of Materials Modeling: Methods}}},\
  \bibinfo {editor} {edited by\ \bibinfo {editor} {\bibfnamefont
  {S.}~\bibnamefont {Yip}}}\ (\bibinfo  {publisher} {Springer Netherlands},\
  \bibinfo {address} {Dordrecht},\ \bibinfo {year} {2005})\ pp.\ \bibinfo
  {pages} {137--148}\BibitemShut {NoStop}%
\bibitem [{\citenamefont {Ho}\ \emph {et~al.}(2008)\citenamefont {Ho},
  \citenamefont {Lignères},\ and\ \citenamefont {Carter}}]{Carter2008}%
  \BibitemOpen
  \bibfield  {author} {\bibinfo {author} {\bibfnamefont {G.~S.}\ \bibnamefont
  {Ho}}, \bibinfo {author} {\bibfnamefont {V.~L.}\ \bibnamefont {Lignères}},\
  and\ \bibinfo {author} {\bibfnamefont {E.~A.}\ \bibnamefont {Carter}},\
  }\href {https://doi.org/https://doi.org/10.1016/j.cpc.2008.07.002} {\bibfield
   {journal} {\bibinfo  {journal} {Computer Physics Communications}\ }\textbf
  {\bibinfo {volume} {179}},\ \bibinfo {pages} {839} (\bibinfo {year}
  {2008})}\BibitemShut {NoStop}%
\bibitem [{\citenamefont {Huang}\ and\ \citenamefont
  {Carter}(2010)}]{HuangCarter2010}%
  \BibitemOpen
  \bibfield  {author} {\bibinfo {author} {\bibfnamefont {C.}~\bibnamefont
  {Huang}}\ and\ \bibinfo {author} {\bibfnamefont {E.~A.}\ \bibnamefont
  {Carter}},\ }\href {https://doi.org/10.1103/PhysRevB.81.045206} {\bibfield
  {journal} {\bibinfo  {journal} {Phys. Rev. B}\ }\textbf {\bibinfo {volume}
  {81}},\ \bibinfo {pages} {045206} (\bibinfo {year} {2010})}\BibitemShut
  {NoStop}%
\bibitem [{\citenamefont {Chen}\ \emph {et~al.}(2013)\citenamefont {Chen},
  \citenamefont {Hung}, \citenamefont {Huang}, \citenamefont {Xia},\ and\
  \citenamefont {Carter}}]{Chen2013}%
  \BibitemOpen
  \bibfield  {author} {\bibinfo {author} {\bibfnamefont {M.}~\bibnamefont
  {Chen}}, \bibinfo {author} {\bibfnamefont {L.}~\bibnamefont {Hung}}, \bibinfo
  {author} {\bibfnamefont {C.}~\bibnamefont {Huang}}, \bibinfo {author}
  {\bibfnamefont {J.}~\bibnamefont {Xia}},\ and\ \bibinfo {author}
  {\bibfnamefont {E.~A.}\ \bibnamefont {Carter}},\ }\href
  {https://doi.org/10.1080/00268976.2013.828379} {\bibfield  {journal}
  {\bibinfo  {journal} {Molecular Physics}\ }\textbf {\bibinfo {volume}
  {111}},\ \bibinfo {pages} {3448} (\bibinfo {year} {2013})}\BibitemShut
  {NoStop}%
\bibitem [{\citenamefont {Baer}\ \emph {et~al.}(2013)\citenamefont {Baer},
  \citenamefont {Neuhauser},\ and\ \citenamefont {Rabani}}]{BaerRabani2013}%
  \BibitemOpen
  \bibfield  {author} {\bibinfo {author} {\bibfnamefont {R.}~\bibnamefont
  {Baer}}, \bibinfo {author} {\bibfnamefont {D.}~\bibnamefont {Neuhauser}},\
  and\ \bibinfo {author} {\bibfnamefont {E.}~\bibnamefont {Rabani}},\ }\href
  {https://doi.org/10.1103/PhysRevLett.111.106402} {\bibfield  {journal}
  {\bibinfo  {journal} {Phys. Rev. Lett.}\ }\textbf {\bibinfo {volume} {111}},\
  \bibinfo {pages} {106402} (\bibinfo {year} {2013})}\BibitemShut {NoStop}%
\bibitem [{\citenamefont {Karasiev}\ \emph {et~al.}(2014)\citenamefont
  {Karasiev}, \citenamefont {Sjostrom},\ and\ \citenamefont
  {Trickey}}]{Karasiev2014}%
  \BibitemOpen
  \bibfield  {author} {\bibinfo {author} {\bibfnamefont {V.~V.}\ \bibnamefont
  {Karasiev}}, \bibinfo {author} {\bibfnamefont {T.}~\bibnamefont {Sjostrom}},\
  and\ \bibinfo {author} {\bibfnamefont {S.}~\bibnamefont {Trickey}},\ }\href
  {https://doi.org/https://doi.org/10.1016/j.cpc.2014.08.023} {\bibfield
  {journal} {\bibinfo  {journal} {Computer Physics Communications}\ }\textbf
  {\bibinfo {volume} {185}},\ \bibinfo {pages} {3240} (\bibinfo {year}
  {2014})}\BibitemShut {NoStop}%
\bibitem [{\citenamefont {Sjostrom}\ and\ \citenamefont
  {Daligault}(2015)}]{Sjostrom2015}%
  \BibitemOpen
  \bibfield  {author} {\bibinfo {author} {\bibfnamefont {T.}~\bibnamefont
  {Sjostrom}}\ and\ \bibinfo {author} {\bibfnamefont {J.}~\bibnamefont
  {Daligault}},\ }\href {https://doi.org/10.1103/PhysRevE.92.063304} {\bibfield
   {journal} {\bibinfo  {journal} {Phys. Rev. E}\ }\textbf {\bibinfo {volume}
  {92}},\ \bibinfo {pages} {063304} (\bibinfo {year} {2015})}\BibitemShut
  {NoStop}%
\bibitem [{\citenamefont {Chen}\ \emph {et~al.}(2015)\citenamefont {Chen},
  \citenamefont {Vella}, \citenamefont {Panagiotopoulos}, \citenamefont
  {Debenedetti}, \citenamefont {Stillinger},\ and\ \citenamefont
  {Carter}}]{Chen2015}%
  \BibitemOpen
  \bibfield  {author} {\bibinfo {author} {\bibfnamefont {M.}~\bibnamefont
  {Chen}}, \bibinfo {author} {\bibfnamefont {J.~R.}\ \bibnamefont {Vella}},
  \bibinfo {author} {\bibfnamefont {A.~Z.}\ \bibnamefont {Panagiotopoulos}},
  \bibinfo {author} {\bibfnamefont {P.~G.}\ \bibnamefont {Debenedetti}},
  \bibinfo {author} {\bibfnamefont {F.~H.}\ \bibnamefont {Stillinger}},\ and\
  \bibinfo {author} {\bibfnamefont {E.~A.}\ \bibnamefont {Carter}},\ }\href
  {https://doi.org/https://doi.org/10.1002/aic.14795} {\bibfield  {journal}
  {\bibinfo  {journal} {AIChE Journal}\ }\textbf {\bibinfo {volume} {61}},\
  \bibinfo {pages} {2841} (\bibinfo {year} {2015})}\BibitemShut {NoStop}%
\bibitem [{\citenamefont {Gao}\ \emph {et~al.}(2015)\citenamefont {Gao},
  \citenamefont {Neuhauser}, \citenamefont {Baer},\ and\ \citenamefont
  {Rabani}}]{Gao2015}%
  \BibitemOpen
  \bibfield  {author} {\bibinfo {author} {\bibfnamefont {Y.}~\bibnamefont
  {Gao}}, \bibinfo {author} {\bibfnamefont {D.}~\bibnamefont {Neuhauser}},
  \bibinfo {author} {\bibfnamefont {R.}~\bibnamefont {Baer}},\ and\ \bibinfo
  {author} {\bibfnamefont {E.}~\bibnamefont {Rabani}},\ }\href
  {https://doi.org/10.1063/1.4905568} {\bibfield  {journal} {\bibinfo
  {journal} {The Journal of Chemical Physics}\ }\textbf {\bibinfo {volume}
  {142}},\ \bibinfo {pages} {034106} (\bibinfo {year} {2015})}\BibitemShut
  {NoStop}%
\bibitem [{\citenamefont {Ding}\ \emph {et~al.}(2018)\citenamefont {Ding},
  \citenamefont {White}, \citenamefont {Hu}, \citenamefont {Certik},\ and\
  \citenamefont {Collins}}]{Ding2018}%
  \BibitemOpen
  \bibfield  {author} {\bibinfo {author} {\bibfnamefont {Y.~H.}\ \bibnamefont
  {Ding}}, \bibinfo {author} {\bibfnamefont {A.~J.}\ \bibnamefont {White}},
  \bibinfo {author} {\bibfnamefont {S.~X.}\ \bibnamefont {Hu}}, \bibinfo
  {author} {\bibfnamefont {O.}~\bibnamefont {Certik}},\ and\ \bibinfo {author}
  {\bibfnamefont {L.~A.}\ \bibnamefont {Collins}},\ }\href
  {https://doi.org/10.1103/PhysRevLett.121.145001} {\bibfield  {journal}
  {\bibinfo  {journal} {Phys. Rev. Lett.}\ }\textbf {\bibinfo {volume} {121}},\
  \bibinfo {pages} {145001} (\bibinfo {year} {2018})}\BibitemShut {NoStop}%
\bibitem [{\citenamefont {White}\ \emph {et~al.}(2018)\citenamefont {White},
  \citenamefont {Certik}, \citenamefont {Ding}, \citenamefont {Hu},\ and\
  \citenamefont {Collins}}]{White2018}%
  \BibitemOpen
  \bibfield  {author} {\bibinfo {author} {\bibfnamefont {A.~J.}\ \bibnamefont
  {White}}, \bibinfo {author} {\bibfnamefont {O.}~\bibnamefont {Certik}},
  \bibinfo {author} {\bibfnamefont {Y.~H.}\ \bibnamefont {Ding}}, \bibinfo
  {author} {\bibfnamefont {S.~X.}\ \bibnamefont {Hu}},\ and\ \bibinfo {author}
  {\bibfnamefont {L.~A.}\ \bibnamefont {Collins}},\ }\href
  {https://doi.org/10.1103/PhysRevB.98.144302} {\bibfield  {journal} {\bibinfo
  {journal} {Phys. Rev. B}\ }\textbf {\bibinfo {volume} {98}},\ \bibinfo
  {pages} {144302} (\bibinfo {year} {2018})}\BibitemShut {NoStop}%
\bibitem [{\citenamefont {Cytter}\ \emph {et~al.}(2018)\citenamefont {Cytter},
  \citenamefont {Rabani}, \citenamefont {Neuhauser},\ and\ \citenamefont
  {Baer}}]{Cytter2018}%
  \BibitemOpen
  \bibfield  {author} {\bibinfo {author} {\bibfnamefont {Y.}~\bibnamefont
  {Cytter}}, \bibinfo {author} {\bibfnamefont {E.}~\bibnamefont {Rabani}},
  \bibinfo {author} {\bibfnamefont {D.}~\bibnamefont {Neuhauser}},\ and\
  \bibinfo {author} {\bibfnamefont {R.}~\bibnamefont {Baer}},\ }\href
  {https://doi.org/10.1103/PhysRevB.97.115207} {\bibfield  {journal} {\bibinfo
  {journal} {Phys. Rev. B}\ }\textbf {\bibinfo {volume} {97}},\ \bibinfo
  {pages} {115207} (\bibinfo {year} {2018})}\BibitemShut {NoStop}%
\bibitem [{\citenamefont {Fabian}\ \emph {et~al.}(2019)\citenamefont {Fabian},
  \citenamefont {Shpiro}, \citenamefont {Rabani}, \citenamefont {Neuhauser},\
  and\ \citenamefont {Baer}}]{Fabian2019}%
  \BibitemOpen
  \bibfield  {author} {\bibinfo {author} {\bibfnamefont {M.~D.}\ \bibnamefont
  {Fabian}}, \bibinfo {author} {\bibfnamefont {B.}~\bibnamefont {Shpiro}},
  \bibinfo {author} {\bibfnamefont {E.}~\bibnamefont {Rabani}}, \bibinfo
  {author} {\bibfnamefont {D.}~\bibnamefont {Neuhauser}},\ and\ \bibinfo
  {author} {\bibfnamefont {R.}~\bibnamefont {Baer}},\ }\href
  {https://doi.org/https://doi.org/10.1002/wcms.1412} {\bibfield  {journal}
  {\bibinfo  {journal} {WIREs Computational Molecular Science}\ }\textbf
  {\bibinfo {volume} {9}},\ \bibinfo {pages} {e1412} (\bibinfo {year}
  {2019})}\BibitemShut {NoStop}%
\bibitem [{\citenamefont {Jiang}\ \emph {et~al.}(2022)\citenamefont {Jiang},
  \citenamefont {Shao},\ and\ \citenamefont {Pavanello}}]{Jiang2022}%
  \BibitemOpen
  \bibfield  {author} {\bibinfo {author} {\bibfnamefont {K.}~\bibnamefont
  {Jiang}}, \bibinfo {author} {\bibfnamefont {X.}~\bibnamefont {Shao}},\ and\
  \bibinfo {author} {\bibfnamefont {M.}~\bibnamefont {Pavanello}},\ }\href
  {https://doi.org/10.1103/PhysRevB.106.115153} {\bibfield  {journal} {\bibinfo
   {journal} {Phys. Rev. B}\ }\textbf {\bibinfo {volume} {106}},\ \bibinfo
  {pages} {115153} (\bibinfo {year} {2022})}\BibitemShut {NoStop}%
\bibitem [{\citenamefont {Fiedler}\ \emph {et~al.}(2022)\citenamefont
  {Fiedler}, \citenamefont {Moldabekov}, \citenamefont {Shao}, \citenamefont
  {Jiang}, \citenamefont {Dornheim}, \citenamefont {Pavanello},\ and\
  \citenamefont {Cangi}}]{Pavanello2022}%
  \BibitemOpen
  \bibfield  {author} {\bibinfo {author} {\bibfnamefont {L.}~\bibnamefont
  {Fiedler}}, \bibinfo {author} {\bibfnamefont {Z.~A.}\ \bibnamefont
  {Moldabekov}}, \bibinfo {author} {\bibfnamefont {X.}~\bibnamefont {Shao}},
  \bibinfo {author} {\bibfnamefont {K.}~\bibnamefont {Jiang}}, \bibinfo
  {author} {\bibfnamefont {T.}~\bibnamefont {Dornheim}}, \bibinfo {author}
  {\bibfnamefont {M.}~\bibnamefont {Pavanello}},\ and\ \bibinfo {author}
  {\bibfnamefont {A.}~\bibnamefont {Cangi}},\ }\href
  {https://doi.org/10.1103/PhysRevResearch.4.043033} {\bibfield  {journal}
  {\bibinfo  {journal} {Phys. Rev. Res.}\ }\textbf {\bibinfo {volume} {4}},\
  \bibinfo {pages} {043033} (\bibinfo {year} {2022})}\BibitemShut {NoStop}%
\bibitem [{\citenamefont {Mi}\ \emph {et~al.}(2023)\citenamefont {Mi},
  \citenamefont {Luo}, \citenamefont {Trickey},\ and\ \citenamefont
  {Pavanello}}]{PavanelloTrickey2023}%
  \BibitemOpen
  \bibfield  {author} {\bibinfo {author} {\bibfnamefont {W.}~\bibnamefont
  {Mi}}, \bibinfo {author} {\bibfnamefont {K.}~\bibnamefont {Luo}}, \bibinfo
  {author} {\bibfnamefont {S.~B.}\ \bibnamefont {Trickey}},\ and\ \bibinfo
  {author} {\bibfnamefont {M.}~\bibnamefont {Pavanello}},\ }\href
  {https://doi.org/10.1021/acs.chemrev.2c00758} {\bibfield  {journal} {\bibinfo
   {journal} {Chemical Reviews}\ }\textbf {\bibinfo {volume} {123}},\ \bibinfo
  {pages} {12039} (\bibinfo {year} {2023})}\BibitemShut {NoStop}%
\end{thebibliography}%



\clearpage
\clearpage 
\setcounter{page}{1}
\renewcommand{\thetable}{S\arabic{table}}  
\setcounter{table}{0}
\renewcommand{\thefigure}{S\arabic{figure}}
\setcounter{figure}{0}
\renewcommand{\thesection}{S\arabic{section}}
\setcounter{section}{0}
\renewcommand{\theequation}{S\arabic{equation}}
\setcounter{equation}{0}
\onecolumngrid

\begin{center}
\textbf{Supplementary Information for\\\vspace{0.2 cm}
\large Mixed Stochastic-Deterministic Density Functional Theoretic Decomposition of Kubo-Greenwood Conductivities in the Projector Augmented Wave Formalism\\\vspace{0.3 cm}}

Vidushi Sharma$^{1,2,3}$, Lee A. Collins$^{1}$, and Alexander J. White$^{1}$

\small

$^1$\textit{Theoretical Division, Los Alamos National Laboratory, Los Alamos, NM 87545, USA}

$^2$\textit{Center for Nonlinear Studies (CNLS), Los Alamos National Laboratory, Los Alamos, NM 87545, USA}

$^3$\textit{Applied Materials and Sustainability Sciences, Princeton Plasma Physics Laboratory, Princeton, NJ 08540-6655, USA}
(Dated: \today)
\end{center}
\section{\label{sec:AppendixA} Real-Time Time-Dependent DFT Dynamics}
\begin{figure}[h]
    \centering
    \includegraphics[width=2.2 in]{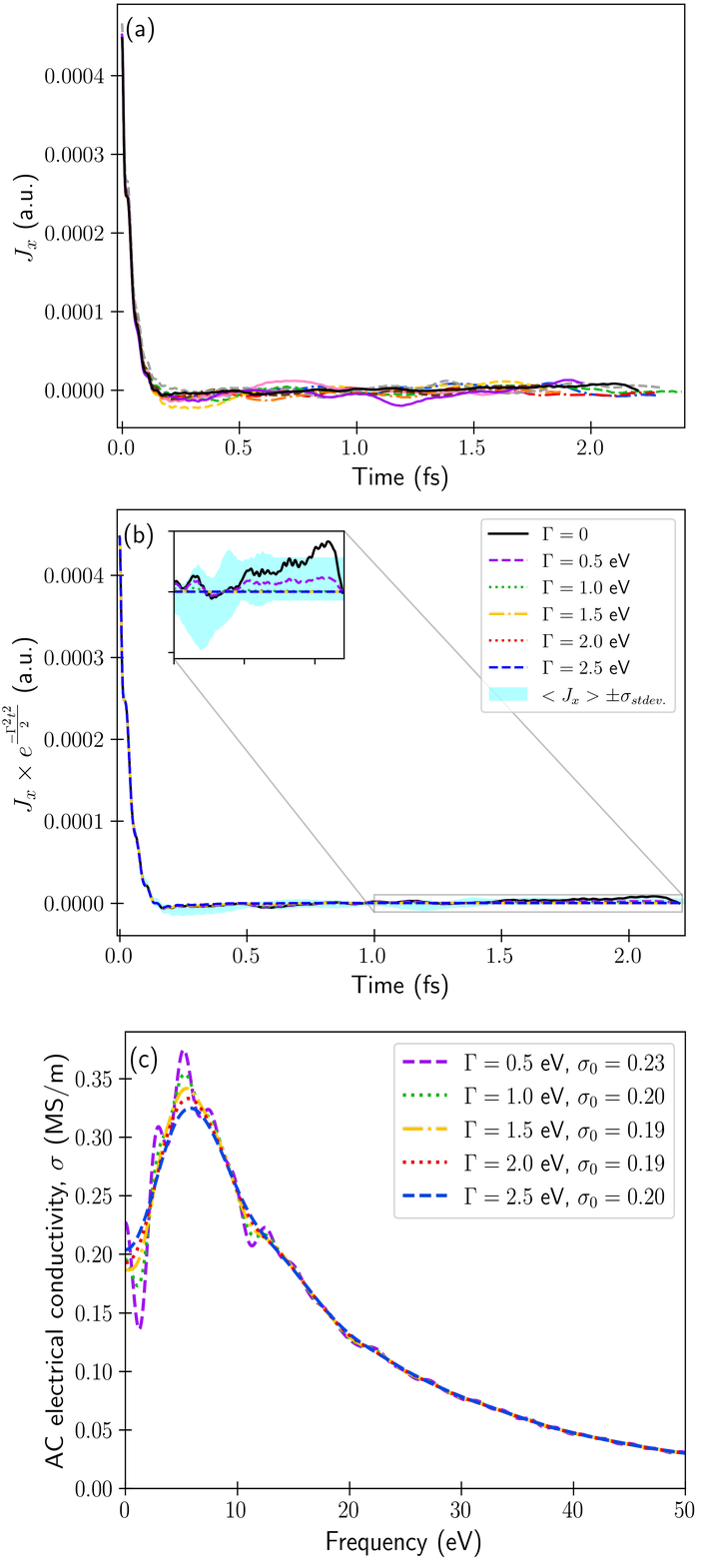}
    \caption{CH/Be at $(\rho, T)= (1.37 \text{ g/cm}^3, 5.0 \text{ eV})$: (a) Current density in the $x-$direction in response to an applied Electric field kick at $\delta(t=0^-)$ computed for ten ionic configurations (shown in different colors) from an MD simulation at thermodynamic equilibrium. The electronic current decays and equilibrates in time as the electronic density evolves according to the TDKS equations. (b) Current density from a sample snapshot (indicated by a black solid line in (a)), multiplied by a filtering function varying with the broadening parameter, $\Gamma$. The cyan-shaded region specifies a standard deviation over different configurations. (c) The AC electrical conductivity along with the DC extrapolation $(\omega\rightarrow0)$ shown for different broadening parameters $(\Gamma)$ for the snapshot selected in (b).}
    \label{fig:SI_current_AC}
\end{figure}

%



\end{document}